 \documentclass[final,3p,times,twocolumn]{elsarticle}

\usepackage{amssymb}

\journal{Astroparticle Physics}

\begin{document}

\begin{frontmatter}



\title{A Measurement of Atomic X-ray Yields in Exotic Atoms and Implications for an Antideuteron-Based Dark Matter Search}


\author[CAL]{T. Aramaki} 
\ead{tsuguo@astro.columbia.edu}

\author[CAL]{S. K. Chan}
\author[LLNL]{W. W. Craig}
\author[LLNL]{L. Fabris} 
\author[CAL]{F. Gahbauer}
\author[CAL]{C. J. Hailey}
\author[CAL]{J. E. Koglin}
\author[LLNL]{N. Madden}
\author[CAL]{K. Mori}
\author[CAL]{H. T. Yu}
\author[LLNL]{K. P. Ziock}

\address[CAL]{Columbia Astrophysics Laboratory, Columbia University, New York, NY 10027, USA}
\address[LLNL]{Lawrence Livermore National Laboratory, Livermore, CA 94550, USA}

\begin{abstract}
The General AntiParticle Spectrometer (GAPS) is a novel approach for the indirect dark matter search that exploits cosmic antideuterons. GAPS utilizes a distinctive detection method using atomic X-rays and charged particles from the exotic atom as well as the timing, stopping range and dE/dX energy deposit of the incoming particle, which provides excellent antideuteron identification. In anticipation of a future balloon experiment, an accelerator test was conducted in 2004 and 2005 at KEK, Japan, in order to prove the concept and to precisely measure the X-ray yields of antiprotonic exotic atoms formed with different target materials \cite{Hailey2006}. The X-ray yields of the exotic atoms with Al and S targets were obtained as $\sim$ 75\%, which are higher than were previously assumed in \cite{Mori2002}. A simple, but comprehensive cascade model has been developed not only to evaluate the measurement results but also to predict the X-ray yields of the exotic atoms formed with any materials in the GAPS instrument. The cascade model is extendable to any kind of exotic atom (any negatively charged cascading particles with any target materials), and it was compared and validated with other experimental data and cascade models for muonic and antiprotonic exotic atoms. The X-ray yields of the antideuteronic exotic atoms are predicted with a simple cascade model and the sensitivity for the GAPS antideuteron search was estimated for the proposed long duration balloon program \cite{Aramaki2013}, which suggests that GAPS has a strong potential to detect antideuterons as a dark matter signature. A GAPS prototype flight (pGAPS) was launched successfully from the JAXA/ISAS balloon facility in Hokkaido, Japan in summer 2012 \cite{Doetinchem2012,Mognet2013} and a proposed GAPS science flight is to fly from Antarctica in the austral summer of 2017-2018.

\end{abstract}

\begin{keyword}
Dark Matter; Antiparticle; antideuteron; Exotic atom; GAPS


\end{keyword}

\end{frontmatter}



\section{Introduction}
\label{Sec:Introduction}

\subsection{Overview}

The General AntiParticle Spectrometer (GAPS) is a novel approach for an indirect dark matter search that exploits cosmic antideuterons. Since the GAPS project utilizes atomic X-rays of exotic atoms to identify antideuterons (see Section \ref{Sec:GAPS}), an accelerator test was conducted in 2004 and 2005 at KEK, Japan, in order to prove the concept and to precisely measure the X-ray yields of antiprotonic exotic atoms formed with different target materials \cite{Hailey2006}. This paper describes not only the detailed analysis for the X-ray yields for antiprotonic exotic atoms (Section \ref{Sec:KEK}), but also the development of a comprehensive cascade model for the exotic atom (Section \ref{Sec:Cascade}). The cascade model was compared and validated with other experimental data and cascade models for muonic and antiprotonic exotic atoms. The results for the accelerator test were used to estimate the X-ray yields for antideuteronic exotic atoms in the GAPS flight experiment. The subsequent GAPS antideuteron sensitivity \cite{Aramaki2013} indicates that the GAPS project has a strong potential to detect antideuterons as a dark matter signature.

\subsection{Dark Matter Candidates}
\label{Sec:DM candidates}

The recent result by the Planck experiment \cite{Planck2013} shows that 68\% of our universe is composed of dark energy, and 27\% is dark matter ($\sim$ 5\% for baryonic matter). The nature and origin of these phenomena, however, are still unknown, and thus are the great cosmological problems of the 21st century. Unlike dark energy, dark matter is well-motivated by many theoretical models, and many experiments are currently being conducted to determine the origin of dark matter.

The existence of dark matter was postulated by Fritz Zwicky in 1933 from the observation of the rotational speed of galaxies. The recent observations of gravitational lensing in the Bullet Cluster (two colliding clusters of galaxies), also indicate the existence of dark matter \cite{Clowe2004}. 

Since dark matter has never been directly observed, it is considered to interact with the Standard Model particles only by the weak force and the gravitational force as seen in rotational curves and gravitational lensing. The small density fluctuations seen in the cosmic microwave background (CMB) \cite{Jarosik2011} and the large scale structure of the present universe indicate that dark matter should be a non-relativistic and massive particle (called cold dark matter). Moreover, it should be stable on a cosmological time scale to be observed in the present universe. Weakly interacting massive particles (WIMPs) are the theoretically best-motivated candidates among the variety of dark matter candidates. Neutralinos, the lightest supersymmetric partner (LSP) in supersymmetric theories, and Kaluza-Klein particles (LKP) and right-handed neutrinos (LZP) in extra dimension theories are examples of popular WIMP candidates.

\subsection{Antideuterons for Dark Matter Search}
\label{Sec:DM search}
There are dozens of experiments designed to search for particles associated with various manifestations of WIMP dark matter categorized into three types, particle collider, direct search, and indirect search. The direct and indirect searches will measure the relic WIMPs, while the particle collider will try to create WIMPs. The direct search measures the recoil energy of a target atom in the detector induced by the interaction with the WIMP, while the indirect search focuses on WIMP-WIMP annihilation products such as electrons, positrons, gamma rays, antiprotons and antideuterons. The detection methods and the background models for each search are different, but also complementary, helping to illuminate the nature of dark matter.

\begin{figure}[!h]
\begin{center} 
\includegraphics*[width=7.5cm]{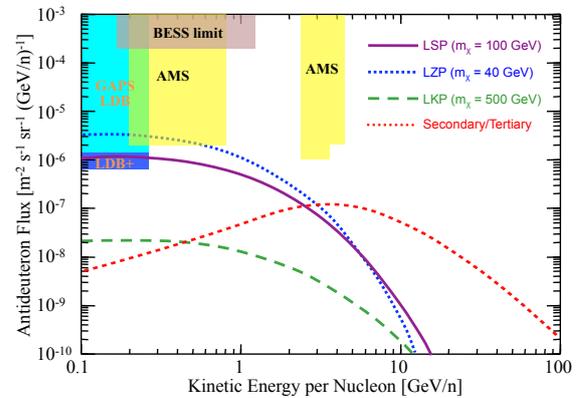}
\end{center}
\caption{Antideuteron flux at the top of the atmosphere, compared with the BESS upper limit \protect \cite{Fuke2005}, and GAPS and AMS sensitivity \protect \cite{Aramaki2013}. The flight altitude for GAPS and BESS is $\sim$ 35-40 km, while AMS is on ISS. The sensitivity for the AMS 5 year flight was estimated, based on \protect \cite{Giovacchini2007}. The blue dashed line (LZP), black dotted line (LSP), and green dot-dashed line (LKP) represent the primary antideuteron fluxes due to the dark matter annihilations \protect \cite{Baer2005}. The red solid line represents the secondary/tertiary flux due to the cosmic-ray interactions \protect \cite{Duperray2005,Salati2010,Ibarra2013}.}
\label{sensitivity}
\end{figure}

Antideuteron production in WIMP-WIMP annihilations was proposed by Donato et al., in 2000 \cite{Donato2000,Donato2008}. The antideuteron flux due to WIMP-WIMP annihilation (called primary flux) can be estimated based on the dark matter density profile of the galaxy, the WIMP-WIMP annihilation channel, the hadronization and coalescence model, and the propagation model. The primary antideuteron flux at the top of the atmosphere due to the WIMP-WIMP annihilation is shown in Figure \ref{sensitivity} (solid purple line: LSP with $m_{\chi}$ $\sim$ 100 GeV, dashed green line: LKP with $m_{\chi}$ $\sim$ 500 GeV, dashed blue line: LZP with $m_{\chi}$ $\sim$ 40 GeV) \cite{Baer2005}. The relatively flat peak is located at $E \sim$ 0.2 GeV/n. The antideuteron flux due to the cosmic-ray interactions with the interstellar medium (secondary/tertiary flux, red dashed line) is also shown in Figure \ref{sensitivity} \cite{Duperray2005,Salati2010,Ibarra2013}. Unlike primary antideuterons, collision kinematics suppress the formation of low-energy secondary antideuterons. Moreover, the interaction rate is drastically decreased at high energy since the flux of the cosmic-ray protons follows the power law, $F_p \sim E^{-2.7}$. Therefore, the primary antideuteron flux is two orders of magnitude larger than the secondary/tertiary antideuteron flux at low energy, and we can clearly distinguish them.

The GAPS and AMS (5 year flight) sensitivities \cite{Aramaki2013}, and the current upper limit for the antideuteron flux obtained by the BESS experiment \cite{Fuke2005} are also shown in Figure \ref{sensitivity}. The flight altitude for GAPS and BESS is $\sim$ 35-40 km ($\sim$ 4-5 g/cm$^2$ atmospheric depth), while AMS is on the International Space Station (ISS). As seen in the figure, the GAPS experiment is more than two order of magnitude more sensitive than the BESS upper limit and 1.5 times more sensitive than the AMS satellite mission. (The sensitivity for a GAPS 210 day flight program (LDB+) is also shown in the figure.) Thus, GAPS has a strong potential to detect antideuterons as the dark matter signature. In the following section, the details of the GAPS project are introduced including the detection concept and the instrumental design. 

\subsection{GAPS Project}
\label{Sec:GAPS}

\subsubsection{Overview of the GAPS Project}

The GAPS project was first proposed in 2002 and was originally named the Gaseous AntiParticle Spectrometer \cite{Mori2002,Hailey2004}. The original GAPS was designed to use a gaseous target, but with further studies, including the KEK (high energy accelerator research organization) beam test in Japan described below, we concluded that a solid target was more efficient and effective for the flight experiment. GAPS is a balloon-borne experiment (flight altitude $\sim$ 35 km), and there are constraints on the size and mass of the payload. Therefore, the solid target can greatly simplify the setup of the GAPS flight module by removing the bulky gas handling system and allowing more complex designs, such as a multi-layer tracker geometry. The higher density of the solid target can also easily slow down and stop more incoming antiparticles, which provides a larger detectable energy range. A GAPS prototype flight (pGAPS) was launched successfully from the JAXA/ISAS balloon facility in Hokkaido, Japan in the summer of 2012 \cite{Doetinchem2012,Mognet2013}, and a proposed GAPS science flight is to fly from Antarctica in the austral summer of 2017-2018.

\subsubsection{Detection Concept}

The GAPS detection method involves capturing antiparticles into a target material with the subsequent formation of an excited exotic atom. A time-of-flight (TOF) system measures the velocity (energy) and direction of an incoming antiparticle. It slows down by the dE/dX energy loss and stops in the target material, forming an excited exotic atom. The exotic atom de-excites in a complex process involving Auger ionization and electron refilling at high quantum number states, followed by the emission of X-rays at the lower quantum states (see Section \ref{Sec:Cascade}). With known atomic number of the target, the Bohr formula for the X-ray energy uniquely determines the mass of the captured antiparticle \cite{Mori2002}. Ultimately, the antiparticle is captured by the nucleus in the atom, where it is annihilated with the emission of pions and protons. The number of pions and protons produced by the nuclear annihilation is approximately proportional to the number of antinucleons, which provides an additional discriminant to identify the incoming antiparticle. The concept of the detection technique has been verified through the accelerator testing at KEK in 2004 and 2005, as described in Section \ref{Sec:KEK}.

\begin{figure}[!h]
\begin{center} 
\includegraphics*[width=7.7cm]{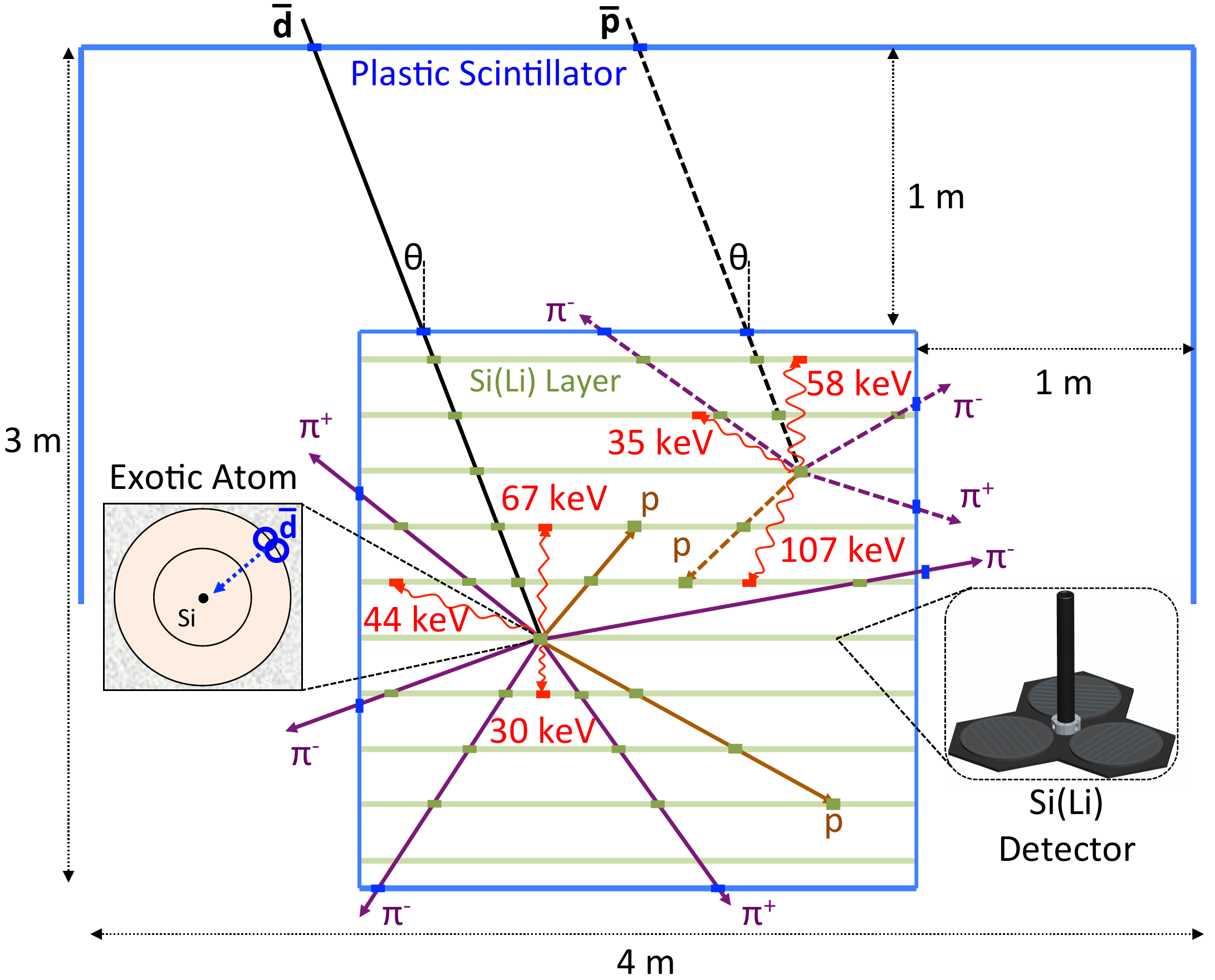}
\end{center}
\caption{The schematic view of the GAPS detector and the detection method. An antiparticle slows down and stops in the Si(Li) target forming an exotic atom. The atomic X-rays will be emitted as it de-excites followed by the pion and proton emission in the nuclear annihilation. The antideuteron identification method from antiprotons is also shown in the schematic view.}
\label{detector}
\end{figure}

Antiprotons are the major background in this experiment, since they can also form exotic atoms and produce atomic X-rays and charged particles. However, the atomic X-rays and the number of pions and protons emitted from the exotic atom uniquely identify the mass of the original antiparticle, as do the depth sensing (stopping range of the incoming particle) and the dE/dX energy loss in each Si(Li) detector, once the velocity of the incoming antiparticle is determined by the TOF system. The three highest antideuteronic X-rays with a Si target in the GAPS detectable energy range are 67 keV, 44 keV and 30 keV, while antiprotonic X-rays are 107 keV, 58 keV, and 35 keV. The number of charged particles produced by the nuclear annihilation for the antideuteronic exotic atom is approximately twice as large as the one for the antiprotonic exotic atom. Additionally, antideuterons with the same speed have a longer stopping range and can go deeper into the detector than antiprotons. Thus, antideuterons with the same stopping range will have a smaller velocity and deposit more energy at each layer than antiprotons, since the dE/dX energy loss is inversely proportional to the velocity squared at low energy. As a result, these detection methods provide an excellent antideuteron identification \cite{Aramaki2013}. The detection concept and the particle identification method in the GAPS project are shown in Figure \ref{detector}.

\subsubsection{Instrumental Design}

The GAPS balloon flight instrument will have a very large, pixellated Si(Li) detector surrounded by a very large TOF system without a pressure vessel as shown in Figure \ref{detector}. There will be 10 layers of detectors surrounded by TOF plastic scintillators, with each layer composed of 4 inch diameter, 2.5 mm thick Si(Li) detectors. Each Si(Li) detector will be segmented into 4 strips, and adjacent tracking layers will have their strips positioned orthogonally, providing modest three-dimensional particle tracking. The tracking geometry can count the number of particles produced in the nuclear annihilation and separately identify atomic X-rays from particle tracks. It also permits direct measurement of particle stopping depth and naturally conforms to the multi-detector geometry. Since each strip is relatively small, $\sim$ 2 cm wide and the layer space is $\sim$ 20 cm, X-rays and charged particles (pions/protons) can be detected separately in the different strips/channels \cite{Hailey2009}. Each Si(Li) layer also works as a degrader and a target material to slow down the incoming antiparticle and to form an exotic atom. Note that since KEK accelerator test focused on the X-ray yields for the antiprotonic exotic atoms, the instrumental setup was different from the one in the GAPS flight instrument as described in Section \ref{Sec:KEK}.

\section{Cascade Model for Exotic Atoms}
\label{Sec:Cascade}

\subsection{Overview of Cascade Model}

As seen in the previous section, X-ray yields of exotic atoms play an important role in the GAPS antideuteron detection. The energy of the atomic X-ray is unique to the exotic atom, allowing us to differentiate antideuterons from other particles, including antiprotons. Therefore, it is crucial to develop a comprehensive cascade model to estimate the X-ray yields for any kind of exotic atom (any negatively charged cascading particles with any target materials) that can form in the GAPS instrument. 

Cascade models for exotic atoms were widely developed after the existence of the exotic atom was predicted in the 1940s. Since the GAPS project focuses on the antiprotonic and antideuteronic exotic atoms formed with a variety of target materials, we have developed a generalized and extendable cascade model. Additionally, since the GAPS detector is designed for X-rays with an energy higher than 10 keV, a very simple cascade model with a few parameters has been developed, focusing on the low $n$ state transitions (E $>$10 keV). The parameters were optimized by the measurement of antiprotonic exotic atoms with Al and S targets at KEK in Japan 2005. The extended cascade model was used to estimate the X-ray yields of the antiprotonic and antideuteronic exotic atom with a Si target and other materials in the GAPS instrument to derive the ultimate antideuteron sensitivity (see Section \ref{Sec:KEK}).

\subsection{Cascade Transitions}

A negatively charged particle ($\mu^-$, $\pi^-$, $K^-$, $\bar{p}$, $\bar{d}$, etc., called ''cascader'' hereafter) will be captured into a target atom at the radius of its outermost electrons after it slows down and its kinetic energy becomes comparable to the binding energy of an electron \cite{Hartmann1990,Gotta2004}. The initial principal quantum number for the exotic atom can be estimated as follows:
\begin{eqnarray*}
n \sim n_e \sqrt{M^*/m^*_e} \ .
\label{initialN}
\end{eqnarray*}
Here, $n_e$ is the principal quantum number of the outermost electron shell of the target atom, $m^*_e$ is the reduced mass of the electron in the target atom and $M^*$ is the reduced mass of the cascader. The cascade model is designed to calculate the probability for the cascader to be in the ($n$, $l$) state, where $l$ is the orbital angular momentum, and to estimate the X-ray yields of the exotic atom as it decays. The cascade model starts at the electron K shell ($n_e$ = 1) and the orbital angular momentum $l$ is assumed to have a statistical distribution, $P_l \propto (2l+1)e^{a l}$. There are ($2l+1$) magnetic quantum numbers, $m = -l+1, -l+2\ ...\ 0\ ...\ l-2, l-1$, for each $l$, and $e^{a l}$ is a correction factor due to the de-excitation at the outer shell, $n_e >$ 1 ($a \sim 0.2$ or less) \cite{Hartmann1990,Gotta2004}. The initial $n$ in the cascade model is about 14 for $\mu^-$, 16 for $\pi^-$, 31 for $K^-$, 42 for $\bar{p}$, and 58 for $\bar{d}$. 

\begin{figure}[!h]
\begin{center} 
\includegraphics*[width=7.5cm]{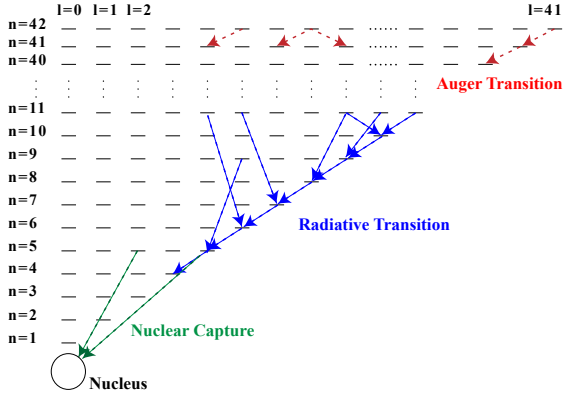}
\end{center}
\caption{The schematic view of the cascade model of the antiprotonic exotic atom. The Auger transitions dominate in high $n$ states, while the radiative transitions dominate in low $n$ states. The nuclear capture takes place in very low $n$ states.}
\label{cascade}
\end{figure}

The three leading de-excitation processes, Auger transition (emission of an Auger electron), radiative transition (emission of an atomic X-ray), and nuclear capture (interaction with the nucleus), dominate the cascade model for atoms with $Z >$ 2, as shown in Figure \ref{cascade}. Auger transitions dominate at the beginning of the cascade, followed by radiative transitions. The nuclear capture takes place in a very low $n$ state. Since the exotic atom can be assumed to be hydrogen-like, the Auger and the radiative transitions with $\Delta l = \pm 1$ dominate due to selection rules \cite{Hartmann1990,Gotta2004}. 

\subsubsection{Auger Transition}
\label{Auger Transition}

In a high $n$ state, an Auger electron is emitted as soon as the energy difference of the initial state ($n_1$, $l_1$) and the final state ($n_2$, $l_2$) exceeds the ionization energy. The Auger transition rate for the K shell and L shell electrons can be estimated by considering the interaction between the cascader and the electron as follows \cite{Eisenberg1961}.

\begin{eqnarray*}
\Gamma^{Aug,K}_{n_1,l_1 \rightarrow n_2,l_2} &=& \frac{32 \pi \alpha c}{a_0 \mu ^2} \left ( \frac{Z^*}{Z} \right ) ^2 \frac{\max (l_1,l_2)}{3(2l_1+1)} \\
&& \cdot \frac{y^2}{1+y^2} \frac{\exp [y(4 \arctan y - \pi)]}{\sinh \pi y} I^2 \\
\Gamma^{Aug,L}_{n_1,l_1 \rightarrow n_2,l_2} &=& \frac{16 \pi \alpha c}{a_0 \mu ^2} \left ( \frac{Z^*}{Z} \right ) ^2 \frac{\max (l_1,l_2)}{3(2l_1+1)} \\
&& \cdot \frac{y^2(4+5y^2)(4+3y^2)}{(4+y^2)^3} \\
&& \cdot \frac{\exp [y(4 \arctan y - \pi)]}{\sinh \pi y} I^2
\end{eqnarray*}
Here, $\mu$, $y$, and $I$ are defined as follows.

\begin{eqnarray*}
\mu &=& M/m_e \\
y &\equiv& \frac{Z^* \alpha}{\sqrt{(T/m_e c^2)^2 + (2T/m_e c^2)}}\\
T &\equiv& \Delta E_{n_1,n_2} - E_{ionization} \\
I^2 &\equiv& \int ^\infty _0 dr\, r^3R(n_1,l_1) R(n_2,l_2) 
\end{eqnarray*}
$\Gamma^{Aug,K}_{n_1,l_1 \rightarrow n_2,l_2}$ ($\Gamma^{Aug,L}_{n_1,l_1 \rightarrow n_2,l_2}$) is the Auger transition rate for emitting K (L) shell electrons with the initial state ($n_1$, $l_1$) and the final state ($n_2$, $l_2$), $a_0$ is the Bohr radius of hydrogen atom, $\alpha$ is the fine structure constant, $Z^*$ is the effective nuclear charge seen from the electron, $T$ is the kinetic energy of the emitted electron, and $R(n,l)$ is the normalized radial function of the exotic atom. The transitions with $\Delta l = \pm 1$ dominate the process, due to the transition selection rules as discussed above. Note that after the electrons are depleted by the Auger transition, the electrons can be filled from adjacent atoms with a refilling rate $\Gamma_{ref}$ and also from the higher shell with the fluorescence rate. The refilling rate can be estimated as follows:

\begin{eqnarray*}
\Gamma_{ref} = n \cdot \sigma \cdot v . 
\end{eqnarray*}
Here, $n$ is the density of target atoms, $\sigma$ is the cross-section for charge transfer ($\sim 10^{-14}$ cm$^2$), and $v$ is the relative velocity of the exotic atom with respect to other atoms of the medium ($< 10^{5}$ cm/s). The typical value of the refilling rate is $\sim 10^{10}$ s$^{-1}$ for low pressure gases and $\sim 10^{13} - 10^{17}$ s$^{-1}$ for solid and metal \cite{Hartmann1990}. 

Since the Auger transition can take place only if an electron occupies a shell state, the time-dependent filling condition of the electron in each shell and the refilling rate from outside, including the electron fluorescence transition (de-excitation) from the outer shell to the inner shell, $\Gamma_{flu}$, needs to be included for a more precise calculation in the cascade model with the time dependent electron population \cite{Koike1996}. However, as described below, this will not affect the X-ray yield in the low $n$ states since the radiative transition rate dominates over the Auger transition rate as $n$ becomes smaller and the radiative transition takes place much faster than the electron refilling rate. Therefore, we simply estimate the modified Auger transition rate, including the electron refilling rate and the fluorescence transition rate, as:

\begin{eqnarray*}
\Gamma^{Aug,K, mod}_{n_1,l_1 \rightarrow n_2,l_2} = \left( \frac{1}{\Gamma^{Aug,K}_{n_1,l_1 \rightarrow n_2,l_2}}+\frac{1}{\Gamma_{ref}} + \frac{1}{\Gamma_{flu}} \right) ^{-1}.
\label{Aug_mod}
\end{eqnarray*}

\subsubsection{Radiative Transition}
\label{Radiative Transition}
The radiative transition rate becomes larger than the Auger process at a relatively low $n$ state. It can be estimated with a perturbation method and in the dipole approximation it follows \cite{Eisenberg1961}.

\begin{eqnarray*}
\Gamma^{Rad}_{n_1,l_1 \rightarrow n_2,l_2} &=& \frac{4e^2}{3\hbar^4 c^3} \left( \frac{a_0}{\mu Z}\right) ^2 \left( \Delta E_{n_1,n_2} \right ) ^3 \\
&& \cdot \frac{\max (l_1,l_2)}{2l_1+1} I^2
\label{radiative}
\end{eqnarray*}
\begin{eqnarray*}
\Delta E_{n_1,n_2} \equiv hcR_y \mu Z^2 \left( \frac{1}{n_1^2} - \frac{1}{n_2^2} \right)
\end{eqnarray*}
Here, $\Gamma^{Rad}_{n_1,l_1 \rightarrow n_2,l_2}$ is the radiative transition rate with the initial state ($n_1$, $l_1$) and the final state ($n_2$, $l_2$), $\Delta E_{n_1,n_2}$ is the energy difference between the initial and final state, and $R_y$ is the Rydberg constant. As seen in the equation, the radiative transition rate increases as $n$ decreases ($\Delta E_{n_1,n_2}$ increases), and becomes the main transition process in low $n$ states. The radiative transitions dominate for $n < 9$ for the antiprotonic exotic atom and $n < 5$ for the muonic exotic atom. Note that the radiative transitions prefer large $\Delta n$ since they are proportional to $\left( \Delta E_{n_1,n_2} \right)^3$, as seen in the equation. However, once the cascader reaches the circular state, ($n, n-1$), the selection rule ($\Delta l = \pm 1$) restricts the transition to ($n, n-1$) $\rightarrow$ ($n-1, n-2$). Therefore, we expect a high X-ray yield in the low $n$ states, since the cascader is predominantly in a circular state at low $n$.

\subsubsection{Nuclear Capture}
\label{Nuclear Capture}
Since the effective Bohr radius for the cascader, $a_0/ \mu$, is much smaller than the Bohr radius, $a_0$, the strong nuclear force interaction between the cascader and the nucleus can become large in low $n$ states. This may terminate the de-excitation cascade of the exotic atom before it reaches the ground state, since the cascader is captured by the nucleus. In particular, the antiproton and the antideuteron annihilate with the nucleus due to the nuclear capture and produce pions and protons. The optical potential between the cascader and the nucleus can be estimated as follows \cite{Batty1981a,Batty1981b}:

\begin{eqnarray*}
U(r) &=& -\frac{2\pi}{M^*} \left( 1+\frac{M^*}{m_N} \right) \bar{a} \rho(r) \\
&\equiv& -(V+iW) \frac{\rho(r)}{\rho(0)} \, \\
&\rho(r)&= \frac{\rho(0)}{1+e^{\frac{r-c}{z}}}.
\end{eqnarray*}
Here, $M^*$ is the reduced mass of the cascader, $m_N$ is the mass of the nucleon, $\bar{a}$ is the average complex ``effective'' hadron-nucleon scattering length (experimentally determined), and $\rho(r)$ is the Fermi distribution with the parameters $\rho(0) = 0.122$ fm$^{-3}$, $c = 1.07 \times A^{1/3}$ fm, and $z = 0.55$ fm \cite{Batty1981a,Batty1981b,Wiegand1969}. 

The nuclear capture rate can be derived with the perturbation method using the imaginary part of the optical potential $W$, as seen below:

\begin{eqnarray*}
\Gamma^{Cap}_{n_1,l_1} &=& \frac{2}{\hbar} \int \mbox{Im}(U(r)) (R(n_1, l_1))^2 r^2 dr \\
&=& \frac{2W}{\hbar} \int \frac{(R(n_1, l_1))^2 r^2}{1+e^{\frac{r-c}{z}}} dr \ .
\end{eqnarray*}
Here, $\hbar$ is the reduced Planck constant and $W$ is $\sim$ 20 MeV (experimentally determined \cite{Batty1981a,Batty1981b,Wiegand1969}). Note that the energy level of the exotic atom might be slightly shifted, due to the strong nuclear force, but the shift is small for low and middle $Z$ atoms and negligible compared with the energy of the atomic X-rays ($\Delta E_{n_1,n_2}$). 

\subsection{Parameter Study and Comparison with Experimental Data}

A Monte Carlo simulation for the cascade model was developed to estimate the X-ray yields of the exotic atom. The simulation takes into account all the possible Auger transitions including the electron refilling and fluorescence transitions, the radiative transitions, and the nuclear capture. It starts at $n_e$ = 1 (electron K shell), and $l$ is determined with the modified statistical distribution $P_l \propto (2l+1)e^{a l}$ as discussed above. Cascaders are then allowed to cascade until they are captured by the nucleus or reach the ($1,0$) state. The absolute X-ray yields, $Y_{n_1 \rightarrow n_2}$, in the low $n$ states (radiative transition dominates) were calculated as follows:

\begin{eqnarray*}
Y_{n_1 \rightarrow n_2} = \sum^{n_1-1}_{l_i = 0} \sum^{n_2-1}_{l_j = 0} \frac{N_{n_1,l_i}}{N_{\mbox{all}}} P^{Rad}_{n_1,l_i \rightarrow n_2,l_j}.
\end{eqnarray*}
Here, the initial and final states are $(n_1, l_1)$ and $(n_2, l_2)$ (no final state for the nuclear capture), $N_{\mbox{all}}$ is the number of antiprotons simulated in the cascade model and $N_{n_1,l_i}$ is the number of antiprotons that cascaded to the state ($n_1,l_i$). 

The Monte Carlo simulation was conducted with three parameters, $a$ for initial angular momentum distribution, $\Gamma_{ref}$ for the electron refilling rate, and $W$ for the optical potential. (The statistical uncertainty was negligible compared to the systematic uncertainty.) Table \ref{Yield_gamma} shows the X-ray yields of antiprotonic exotic atoms (Al target) with the different values of $\Gamma_{ref}$ around the empirical values, $\Gamma_{ref} = 10^{16}$ s$^{-1}$ ($a = 0.16$ and $W$ = 10 MeV). This indicates, as discussed above, the X-ray yields at low $n$ states were not affected by the electron refilling rate. The results are also consistent with models including the time dependent electron population\footnote{private communication with Dr. Takahisa Koike (RIKEN Japan)}. Table \ref{Yield_W} shows the X-ray yields of antiprotonic exotic atoms (Al target) with the different values of $W$ ($a = 0.16$ and $\Gamma_{ref} = 10^{16}$ s$^{-1}$). This also indicates that $W$ affects only one transition, the lowest $n$, as expected. Table \ref{Yield_alpha} shows the X-ray yields of antiprotonic exotic atoms (Al target) with the different values of $a$ ($W$ = 10 MeV and $\Gamma_{ref} = 10^{16}$ s$^{-1}$). As seen in the tables, the X-ray yields are driven mainly by $a$, the initial angular momentum distribution, except the last transition is strongly affected by $W$, the nuclear potential. 

\begin{table}[hbtp]
\centering
\begin{tabular}{c|c|c|c|c}
$\Gamma_{ref}$ [s$^{-1}$] &  $10^{13}$ & $10^{14}$ & $10^{16}$ & $10^{18}$\\
\hline
92 keV ($5 \rightarrow 4$)& 72\% & 68\% & 67\% & 67\% \\
\hline
50 keV ($6 \rightarrow 5$) & 91\% & 84\% & 82\% & 81\% \\
\hline
30 keV ($7 \rightarrow 6$) & 83\% & 71\% & 69\% & 68\% \\
\end{tabular}
\caption{X-ray yields of the antiprotonic exotic atom (Al target) with different values of $\Gamma$ ($a = 0.16$, $W$ = 10 MeV).}
\label{Yield_gamma}
\end{table}

\begin{table}[hbtp]
\centering
\begin{tabular}{c|c|c|c|c|c}
$W$ [MeV] &  0 & 10 & 30 & 50 & 100\\
\hline
92 keV ($5 \rightarrow 4$)& 89\% & 67\% & 46\% & 35\% & 22\% \\
\hline
50 keV ($6 \rightarrow 5$) & 82\% & 82\% & 82\% & 81\% & 82\% \\
\hline
30 keV ($7 \rightarrow 6$) & 69\% & 69\% & 69\% & 68\%& 69\% \\
\end{tabular}
\caption{X-ray yields of the antiprotonic exotic atom (Al target) with different values of $W$ ($a = 0.16$, $\Gamma_{ref} = 10^{16}$ s$^{-1}$).}
\label{Yield_W}
\end{table}

\begin{table}[hbtp]
\centering
\begin{tabular}{c|c|c|c|c}
$a$ & 0 & 0.08 & 0.16 & 0.24\\
\hline
92 keV ($5 \rightarrow 4$)& 41\% & 58\% & 67\% & 71\% \\
\hline
50 keV ($6 \rightarrow 5$) & 46\% & 69\% & 82\% & 88\% \\
\hline
30 keV ($7 \rightarrow 6$) & 37\% & 56\% & 69\% & 75\% \\
\end{tabular}
\caption{X-ray yields of the antiprotonic exotic atom (Al target) with different values of $a$ ($W$ = 10 MeV, $\Gamma_{ref} = 10^{16}$ s$^{-1}$).}
\label{Yield_alpha}
\end{table}

Additionally, our cascade model was compared with the data for muonic exotic atoms, which are widely measured in experiments \cite{measday2001}. The nuclear absorptions are not seen in the muonic exotic atoms (except for high Z targets) and therefore, there is only one parameter, $a$, to control the X-ray yields at low $n$ states. Table \ref{Yield_mu} shows the comparison with the experimental data and a cascade model developed by Vogel and Hartmann for the muonic exotic atoms \cite{measday2001,Vogel1980,Hartmann1982}. The parameters used here were $W = 0$ MeV (no nuclear absorption), $\Gamma_{ref} = 10^{15}$ s$^{-1}$, and $a = $ 0.16, -0.18, -0.01 for Al, Fe and Au targets, obtained by the empirical fit. 

\begin{table}[hbtp]
\centering
\begin{tabular}{c|c|c|c}
Transition & exp & our model & model in \cite{Vogel1980,Hartmann1982}\\
\hline
Al ($2 \rightarrow 1$)& 80\% & 78\% & 80\% \\
\hline
\ \ \ \ \ ($3 \rightarrow 2$) & 63\% & 60\% & 60\% \\
\hline
\ \ \ \ \ ($4 \rightarrow 3$) & 34\% & 38\% & 42\% \\
\hline
\hline
Fe ($2 \rightarrow 1$)& 72\% & 71\% & 74\% \\
\hline
\ \ \ \ \ ($3 \rightarrow 2$) & 44\% & 49\% & 45\% \\
\hline
\ \ \ \ \ ($4 \rightarrow 3$) & 33\% & 33\% & 33\% \\
\hline
\hline
Au ($2 \rightarrow 1$)& 90\% & 94\% & 95\% \\
\hline
\ \ \ \ \ ($3 \rightarrow 2$) & 80\% & 85\% & 84\% \\
\hline
\ \ \ \ \ ($4 \rightarrow 3$) & 76\% & 75\% & 76\% \\
\end{tabular}
\caption{Experimental data and cascade models for X-ray yields of the muonic exotic atoms with Al, Fe and Au targets}
\label{Yield_mu}
\end{table}

As seen above, X-ray yields at low $n$ states in our cascade model are in good agreement with both experimental data (muonic exotic atoms) and other cascade models (both muonic and antiprotonic exotic atoms) with a time dependent electron population.

\section{Accelerator Test at KEK}
\label{Sec:KEK}
\subsection{Overview of Accelerator Test}

The KEK facility is located north of Tokyo, in Tsukuba, Japan. During the course of the experiments the proton synchrotron produced an 8 GeV (up to 12 GeV) proton beam in the main ring. The H$^{-}$ ion source generated in the plasma chamber was injected into the pre-injector, followed by the linac, booster synchrotron and main ring and accelerated to 750 keV, 40 MeV, 500 MeV and 8 GeV, respectively. Our experiment was performed at the $\pi$2 secondary beam line, which delivers copious particles including antiprotons generated by the proton beam hitting an internal target in the main ring\footnote{KEK PS experiment [http://www-ps.kek.jp/kekps/index.html]}. 

The beam test was conducted at KEK in 2004 and 2005 to verify the GAPS original concept described in \cite{Mori2002} and measure the X-ray yields of the antiprotonic exotic atom with several different target materials. The results constrained the parameters in the cascade model described in Section \ref{Sec:Cascade}. This also allowed us to extend the cascade model to any exotic atoms and estimate the X-ray yield of the antiprotonic and antideuteronic exotic atoms in the GAPS experiment. 

\subsection{Experimental Setup}

\begin{figure}[!h]
\begin{center} 
\includegraphics*[width=7.5cm]{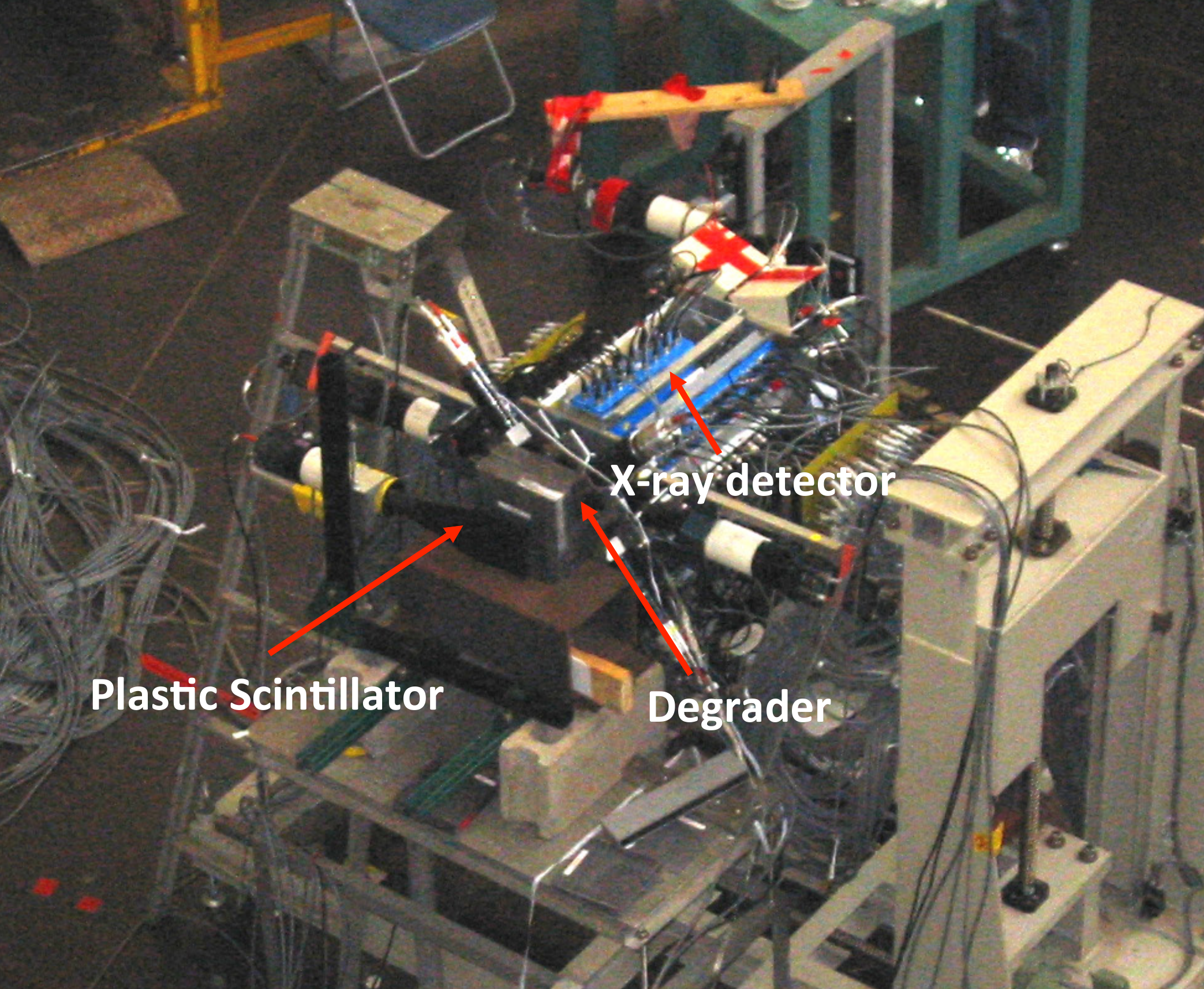}
\end{center}
\caption{KEK experimental setup.}
\label{KEK_pic}
\end{figure}

The experimental setup in the KEK test was composed of a TOF, degraders (lead brick and sheets), shower counters, a target and X-ray detectors. The antiprotons in the beam were first identified by the TOF system, since antiprotons are slower than the other particles in the beam. The degrader slowed down antiprotons and stopped them in the target material where they formed an excited antiprotonic exotic atom. Atomic X-rays and charged particles are emitted in the decay of the exotic atom as discussed in the previous chapter. A Sodium Iodide doped with Thallium, Nal(Tl), detector array was installed around the target material and detected the atomic X-rays and pions. The shower counters monitored the energy deposited by the particles in the beam and distinguished antiprotons from other particles, including the in-flight annihilation products. 

\begin{figure}[!h]
\begin{center} 
\includegraphics*[width=7.5cm]{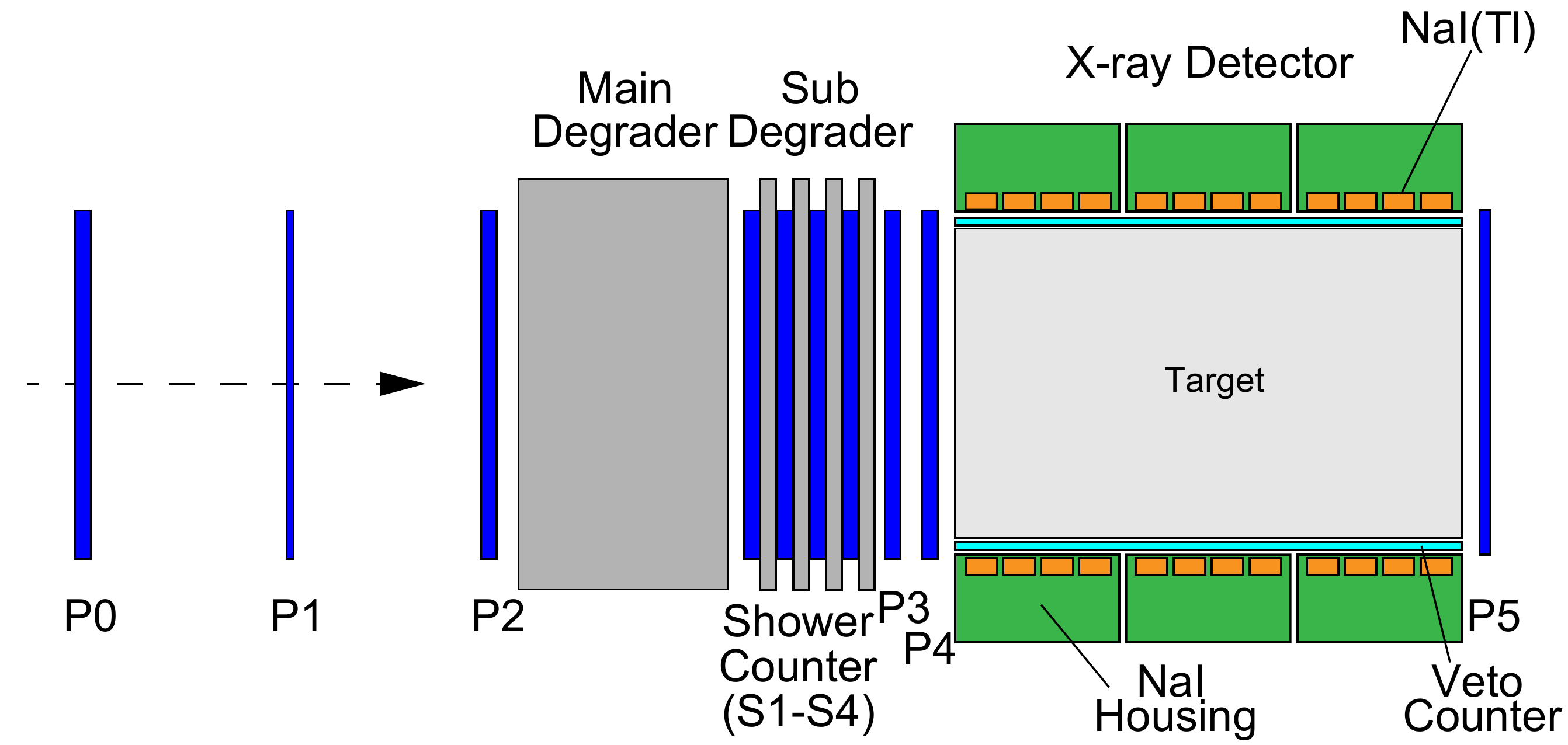}
\end{center}
\caption{The schematic view of the experimental setup at KEK in 2005. It was composed of a TOF system (P0-P5), degraders (lead brick and sheets), shower counters (S1-S4), a target and X-ray detectors. The distance between the P0 and P2 counters is 6.5 m and the overall length of the X-ray detector is $\sim$ 50 cm.}
\label{KEK05}
\end{figure}

\begin{figure}[!h]
\begin{center} 
\includegraphics*[width=7.5cm]{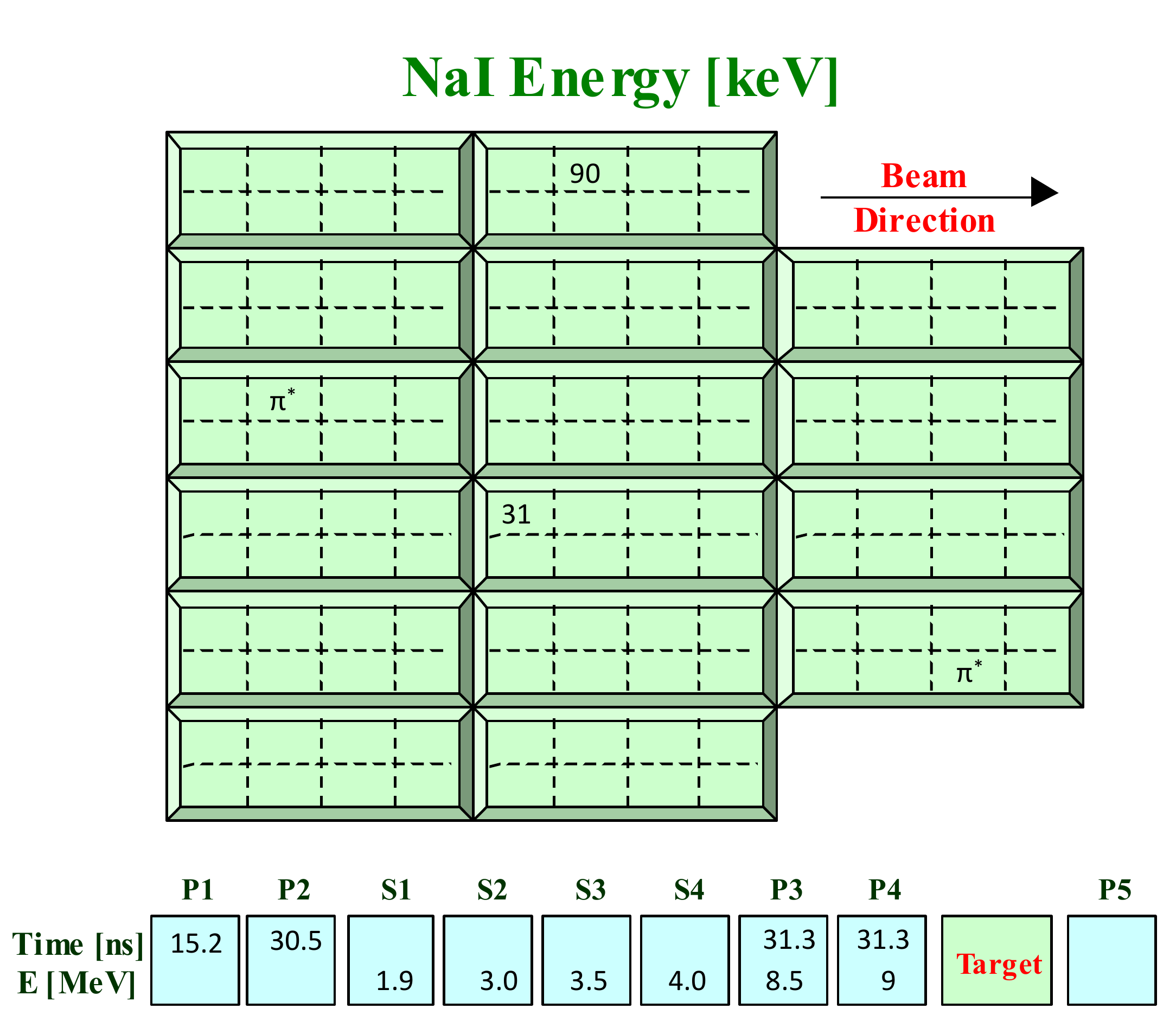}
\end{center}
\caption{A typical antiproton event for the Al target. Numbers are the energy of the stopped X-rays and $\pi^{*}$ indicates a pion hit.}
\label{NaI}
\end{figure}

While gaseous targets and a few liquid targets were used in 2004, liquid and solid targets were tested in 2005, since they are simpler to implement in the realistic design for the balloon experiment. The actual picture and the schematic view of the experimental setup in 2005 are shown in Figures \ref{KEK_pic} and \ref{KEK05}. Figure \ref{NaI} shows an unfolded view of the cylindrical detector module and a typical stopped antiproton event for the Al target. Numbers are the energy of the stopped X-rays and $\pi^{*}$ indicates a pion hit.

\subsubsection{Beam Profile}

The momentum of the beam was controlled and focused by dipole and quadrupole magnets, while the momentum spread was controlled by a shutter. The particles were delivered in 1.5 s long spills, and each spill was separated by a 4 s interval. A momentum of 1 GeV/c was used in all of the 2005 measurements. Up to several GeV/c the antiproton flux from the $\pi$2 beam line increases with increasing momentum; however, losses due to annihilation in the degrader increase for the thicker degraders required to stop higher momentum antiprotons. Around 1 GeV/c was found to provide the highest rate of antiproton stops in our target during our 2004 measurements. The beam spill with a momentum of 1 GeV/c contained about 20-30 antiprotons, 10$^{5}$ $\pi^{-}$, and a somewhat smaller number of K$^{-}$ and e$^{-}$, as measured in the 2004 experiment, and these numbers were consistent with the data sheets provided by KEK.

The spatial beam profile at the P0 counter was measured by changing the last dipole magnet, which controlled the horizontal direction, and the height of the remote controlled table (for the vertical direction). The measured beam profiles at P0, P1 and P2 were used as input in the GEANT4 simulation \footnote{GEometry ANd Tracking, a toolkit for the simulation of the passage of particles through matter, developed by CERN.} together with a TURTLE beam line optics ray trace \footnote{PSI Graphic Turtle Framework by U. Rohrer based on a CERN-SLAC-FERMILAB version by K.L. Brown et al.} to simulate the beam profile, divergence and momentum bite.

\subsubsection{Time of Flight System}

The TOF system in the KEK 2005 test was composed of 6 scintillation counters, P0-P5. The TOF timing, the travel time of the incoming particle between the P0 and Pi (i = 1, 2, 3, 4, 5) counters, allowed us to identify the incoming particle, since all the particles in the beam had a fixed momentum and the antiprotons were much slower than the other lighter particles (see below). The P0, P2, P3 and P4 counters had a dimension of 12 cm $\times$ 12 cm and a thickness of 1.0 cm, while the P1 and P5 counters had a thickness of 0.2 cm. The paddles were coupled to the light guide and then to the 2 inch fast photomultiplier tube (Photonis XP2020). A high voltage of $\sim$ -1800V was applied to the PMT base (Photonis S5632).

The P0, P1, and P2 counters were used for timing only, while the P3, P4 and P5 counters were used for both timing and energy deposition. The timing at each counter was measured relative to the accelerator beam structure by passing the signal from the last diode through a fast timing preamplifier (Ortec VT120b), followed by a constant fraction discriminator. The time of flight (TOF) between the P0 counter and the P1-P5 counters were measured using time- to-analog converters (TAC, Canberra 2020). The dE/dX energy deposit was characterized using the signal from the PMT anode passed through a preamplifier (Camberra 2005) followed by a spectroscopy amplifier (Ortec 452).

\subsubsection{Shower Counter}

The four shower counters, S1-S4, were installed behind the main degrader in 2005 (see Figure \ref{KEK05}), and each of them had a dimension of 12 cm $\times$ 12 cm $\times$ 0.5 cm. A 0.25 inch lead sheet was sandwiched between every pair of counters to slow down the incoming particle. Each counter was coupled to the light guide and PMT (Photonis 2042 and 2072 PMT). The shower counter allowed us to distinguish antiprotons from other particles by measuring the dE/dX energy loss, since non-relativistic slow antiprotons deposit more energy than relativistic particles such as $\pi^{-}$.

The veto counters (6 cm wide, 1 mm thick ribbon scintillation fibers, coupled to a Hamamatsu R1942A 1 inch PMT) were installed between the target and the X-ray detectors. They were designed to monitor the off-axis antiprotons hitting the detector and the frame without stopping in the target material. However, since the energy resolution of these counters was relatively coarse, it was difficult to uniquely identify potential off-axis antiproton interactions from annihilation products produced in the target.

\subsubsection{X-ray Detector}

The X-ray detectors were 128 NaI(TI) crystals (1 inch $\times$ 1 inch $\times$ 5 mm). The NaI(TI) detector emits scintillation light proportional to the deposited energy, $\sim$ 40 photons/keV. Since the NaI(Tl) is a relatively high Z material, up to 300 keV X-rays (20 keV threshold) can be photo-absorbed in the 5 mm thick crystal. Each crystal is coupled to a Hamamatsu 1 inch PMT (R1924A) on the back surface. The wavelength of the scintillation light is $\sim$ 410 nm, where the quantum efficiency of the PMT has a peak. Every 8 crystals and PMTs, separated from each other by 1.5 inch, are mounted in a tightly sealed steel housing with a 0.125 mm Al window. Each PMT is connected to the custom made PMT base and $\sim$ -800V HV was applied. The preamplifier was mounted inside the housing and the gain for each detector was controlled externally. Sixteen sets of detectors were mounted around the target as seen in Figure \ref{NaI}.

\subsubsection{Target Material}

In 2005, four target materials were chosen based on the energy of the atomic X-rays in their antiprotonic exotic atom, which needed to be in the useful energy range of the X-ray detector, 25 keV $<$ E $<$ 300 keV. The detectable antiprotonic atomic X-rays for each target tested in KEK are shown in Table \ref{target}. 

\begin{table}[hbtp]
\caption{Antiprotonic atomic X-rays for each target (25 keV $<$ E $<$ 300 keV)}
\begin{center}
\small\addtolength{\tabcolsep}{-4pt}
\begin{tabular}{c|c|c|c|c|c|c}
Target & X$_{1}$ & X$_{2}$ & X$_{3}$ & X$_{4}$ & X$_{5}$ & X$_{6}$ \\
\hline
Al & 92 keV & 50 keV & 30 keV & - & - & - \\
\hline
S & 139 keV & 76 keV & 46 keV & 30 keV & - & - \\
\hline
Cl & 86 keV & 52 keV & 34 keV & 23 keV & - & - \\
\hline
Br & 145 keV & 99 keV & 71 keV & 52 keV & 41 keV & 31 keV\\ 
\end{tabular}
\end{center}
\label{target}
\end{table}%

\begin{figure}[!h]
\begin{center} 
\includegraphics*[width=7.5cm]{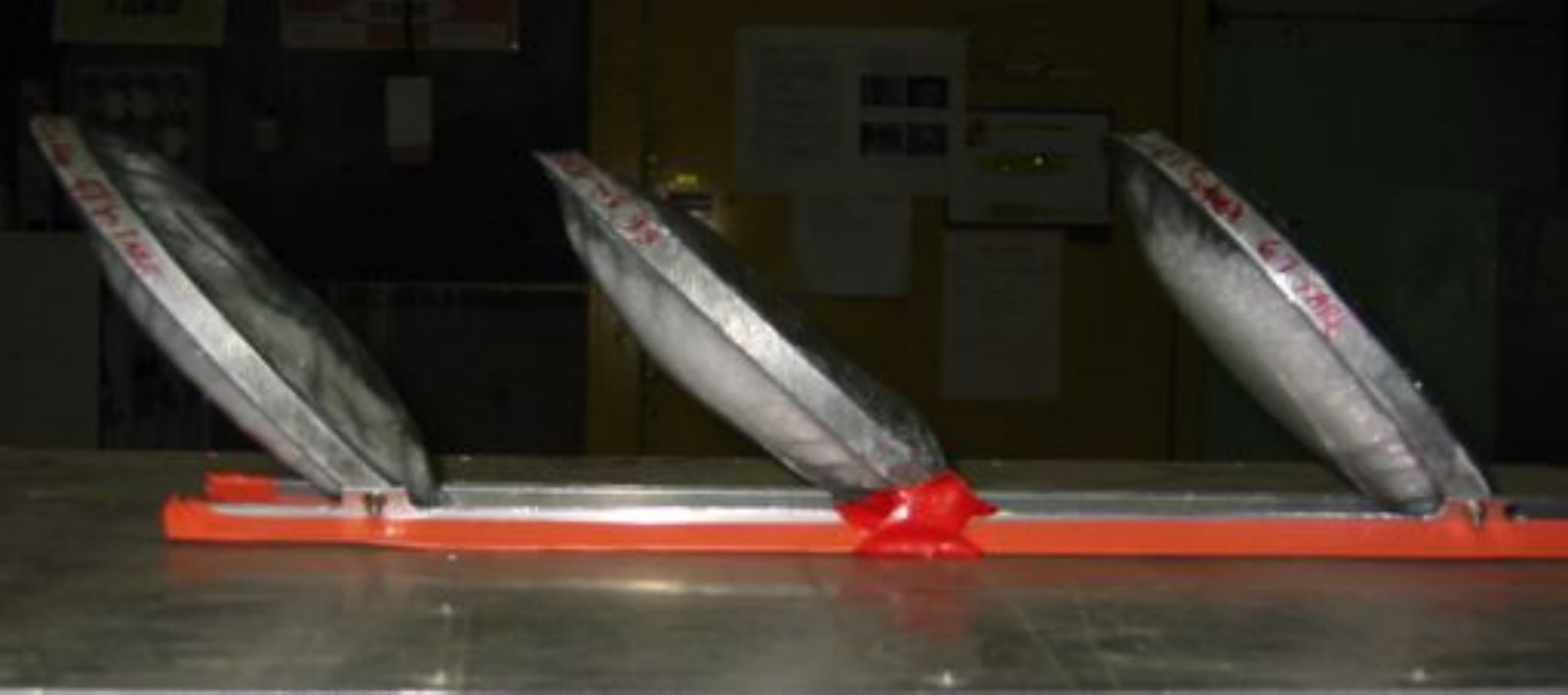}
\end{center}
\caption{Sulfur target geometry}
\label{target_holder}
\end{figure}

Al (Aluminum wool), S (Sulfur), CBr$_{4}$ (Tetrabromomethane) and CCl$_{4}$ (Carbon tetrachloride) targets were tested in 2005 and we will focus on the Al (Z = 13) and S (Z = 16) targets in this paper to estimate the X-ray yields for the Si (Z = 14) target for the GAPS balloon experiment. It is also more challenging to analyze the data for the CBr$_4$ and CCl$_4$ targets since they are compounds and many atomic X-rays can be produced in the small energy region. The Al wool was filled into two 1 mm thick plastic bottles, each with a diameter of 12 cm and 22 cm in length, and the average density was $\sim$ 0.111g/cm$^3$. The target holder for the Sulfur powder was framed with Al pipes of diameter 12 cm cut at a 45 degree angle, and both openings were covered with 1 mm thick plastic sheets (see Figure \ref{target_holder}). This is the most favorable geometry for X-rays to escape in the cylindrical geometry. The holders were placed onto two guided rails to minimize the blockage of X-rays from the target. 

\subsubsection{Degrader Thickness and Range Curve}

\begin{figure}[!b]
\begin{center} 
\includegraphics*[width=7.5cm]{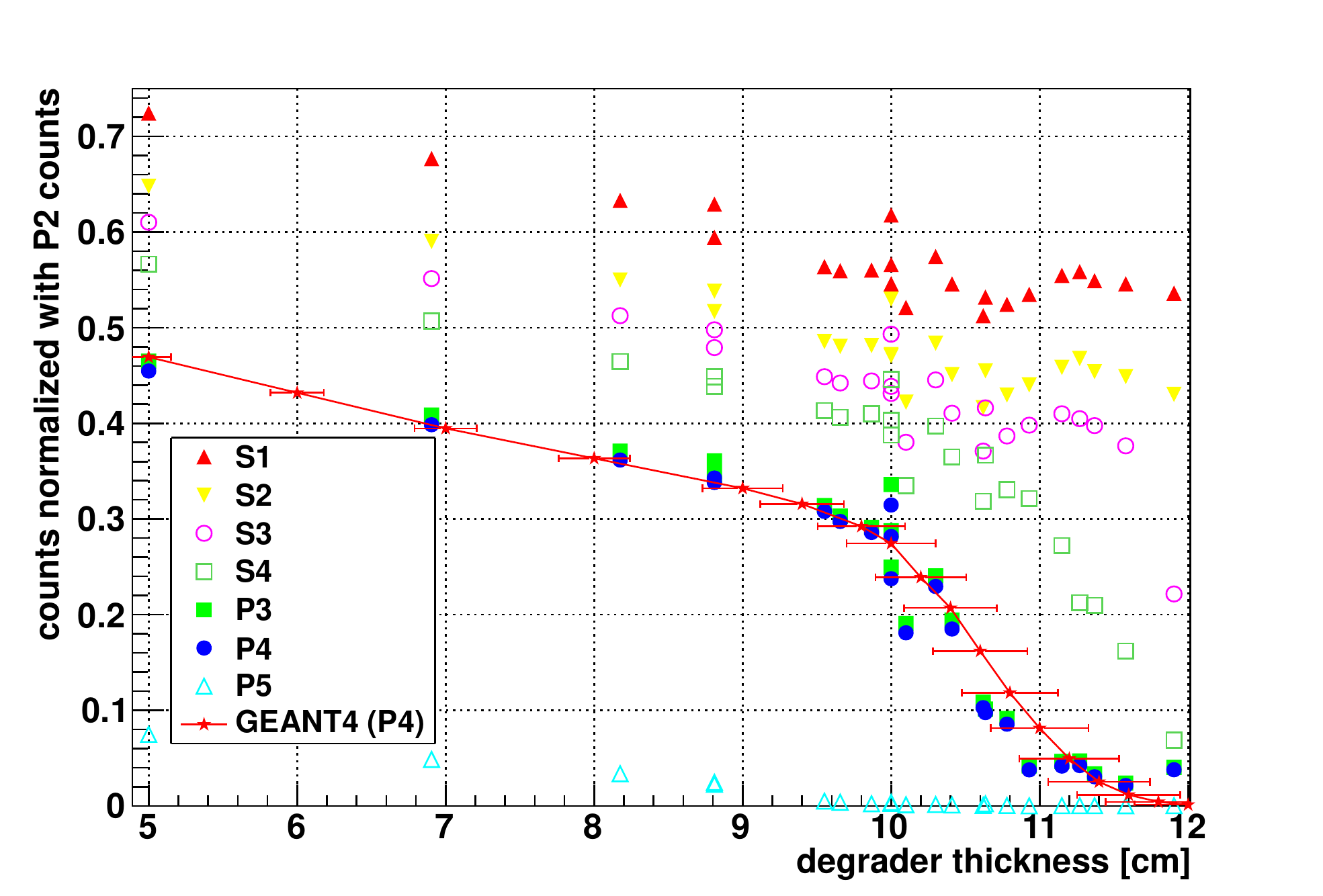}
\end{center}
\caption{Number of hits at each counter vs. degrader thickness (cm). The counts were normalized to the counts at the P2 counter.}
\label{RangeCurve}
\end{figure}

Since antiprotons in the beam were too energetic to stop in the target, a combination of active and passive degraders were used to slow down the antiprotons before they entered the GAPS target region (see Figure \ref{KEK05}). The optimized total thickness of degrader was estimated by measuring the number of events at the P4 counter (just before the target) with different thicknesses of degrader. Since the number of antiprotons in the beam was very small, in order to have better statistics we used positively charged beam (protons and $\pi^{+}$) with the same magnet settings for the beam except for the polarity. Figure \ref{RangeCurve} shows the number of protons at each counter, normalized with the number of protons at the P2 counter. The GEANT4 simulation result at the P4 counter is also shown in the figure, taking into account the uncertainty on the thickness and density for each lead brick and sheet ($\sim$ 3\%). The data and the simulation result are in good agreement with each other and the number of protons at the P4 counter rapidly decreased as the thickness of the degrader increased to $\sim$ 10.3 cm. This implies that there were many slow protons present at the P4 counter with this thickness and thus we decided to use a total thickness of the degrader of 10.3 cm in the experiment.

\subsection{Antiproton Selection}

Since the momentum of the beam was set as 1 GeV/c by the dipole and quadrupole magnets, the antiprotons in the beam can be distinguished from other particles by the velocity, $\beta$. The TOF timing and energy deposit in the plastic scintillator for antiprotons are larger than other particles in the beam (mainly pions) since $\beta$ for antiprotons is smaller and dE/dx energy loss is proportional to $\sim \beta^{-2}$. 

\begin{figure}[!h]
\begin{center} 
\includegraphics*[width=8.0cm]{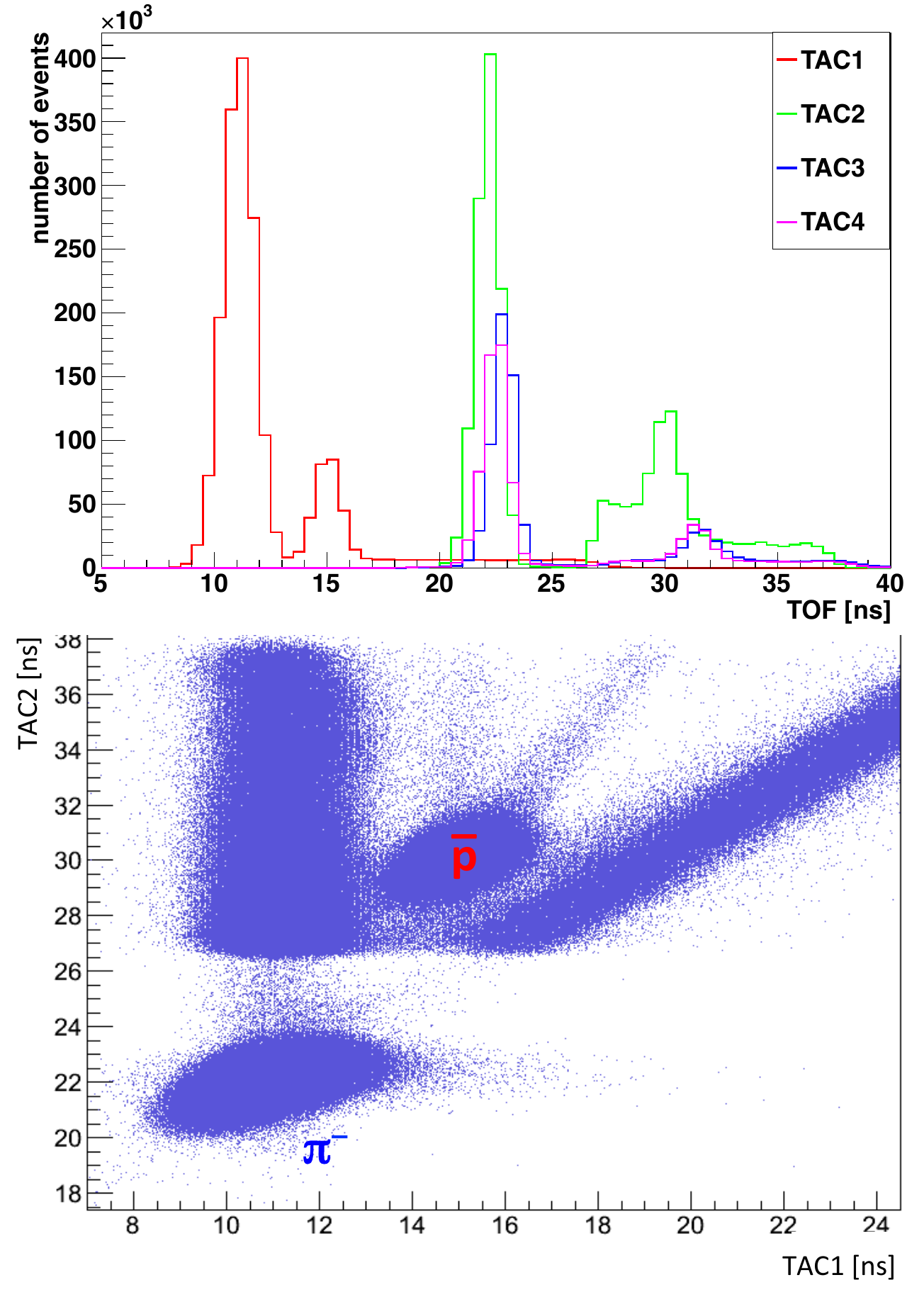}
\end{center}
\caption{TOF timing at TAC1 (red), TAC2 (green), TAC3 (blue), TAC4 (purple), and TAC1 vs. TAC2 (bottom).}
\label{KEK_TAC}
\end{figure}

Figure \ref{KEK_TAC} shows the TOF timing at TAC1 (red), TAC2 (green), TAC3 (blue) and TAC4 (purple), between the P0 and Pi (i = 1, 2, 3, 4) counters. A plot for TAC1 vs. TAC2 was also shown below. Two peaks, relativistic pions (pre-scaled) and antiprotons, are seen at each plot. The selected cuts, peak $\pm$ 1 ns, were applied for each TOF timing to select antiproton events. Since the P5 counter was placed $\sim$ 60 cm away from the P4 counter, the cut on the TAC5 was set as ``(TAC4 lower limit + 2 ns) or no hit''. Tables \ref{TOFcut} shows the applied cuts for each TOF timing. 

\begin{table}[hbtp]
\centering
\begin{tabular}{c|c|c}
 & lower limit & upper limit \\
\hline
TAC1 & 14.0 ns & 16.0 ns \\
\hline
TAC2 & 29.0 ns & 31.0 ns \\
\hline
TAC3 & 30.5 ns & 32.5 ns \\
\hline
TAC4 & 30.5 ns & 32.5 ns \\
\hline
TAC5 & \multicolumn{2}{|c}{$>$ 32.5 ns or no hits} \\ 
\end{tabular}
\caption{Antiproton selection cuts on each TOF timing}
\label{TOFcut}
\end{table}

\begin{figure}[!b]
\begin{center} 
\includegraphics*[width=8.0cm]{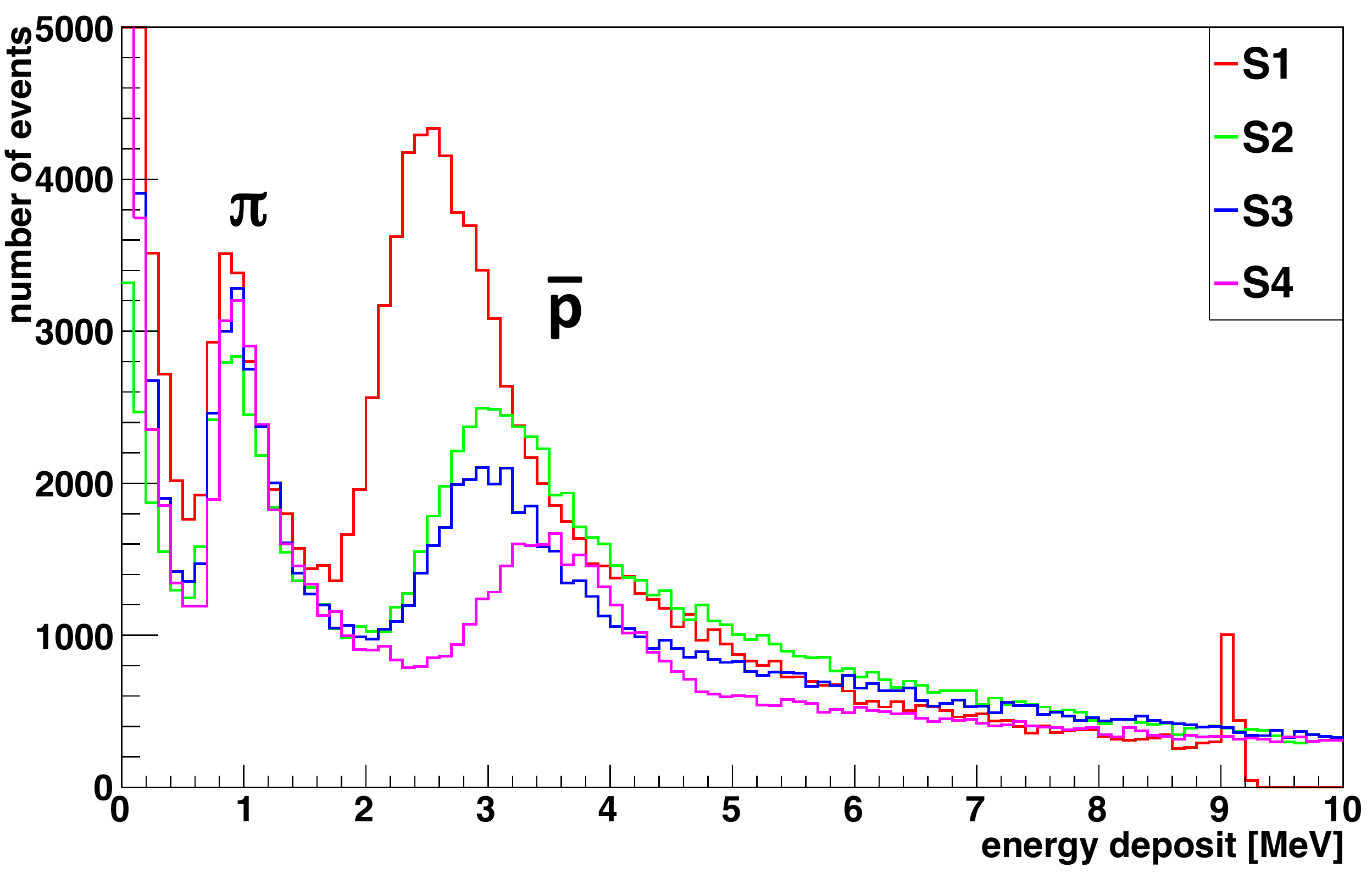}
\end{center}
\caption{dE/dX energy deposit in the S1 (red), S2 (green), S3 (blue) and S4 (purple) counters.}
\label{KEK_S}
\end{figure}

Figure \ref{KEK_S} shows the dE/dX energy deposit in the shower counters (S1: red, S2: green, S3: blue, S4: purple). Energy was calibrated with the relativistic pions in the beam and the GEANT4 simulation with the actual beam profile. Two peaks, relativistic pions (pre-scaled) and antiprotons, are also seen at each plot. The antiproton selection cuts were applied for each dE/dX energy deposit. The applied cuts for each dE/dX energy deposit are shown in Table \ref{dE/dXcut}. Note that we were not able to set the upper limits on the cuts for the P3 and P4 dE/dX energy deposits since the signals were saturated at E $>$ 10 MeV. After applying the cuts, the X-ray lines are seen in the spectrum (see Fig \ref{Al_fit_KEK}).

\begin{table}[hbtp]
\centering
\begin{tabular}{c|c|c}
 & lower limit & upper limit \\
\hline
S1 & 1.8 MeV & 3.2 MeV \\
\hline
S2 & 2.2 MeV & 4.2 MeV \\
\hline
S3 & 2.2 MeV & 4.2 MeV \\
\hline
S4 & 2.6 MeV & 5.0 MeV \\
\hline
P3 & 8.0 MeV & - \\
\hline
P4 & 8.0 MeV & - \\
\end{tabular}
\caption{Antiproton selection cuts on each dE/dX energy deposit}
\label{dE/dXcut}
\end{table}

\subsection{Background Model}

\begin{figure}[!b]
\begin{center} 
\includegraphics*[width=7.5cm]{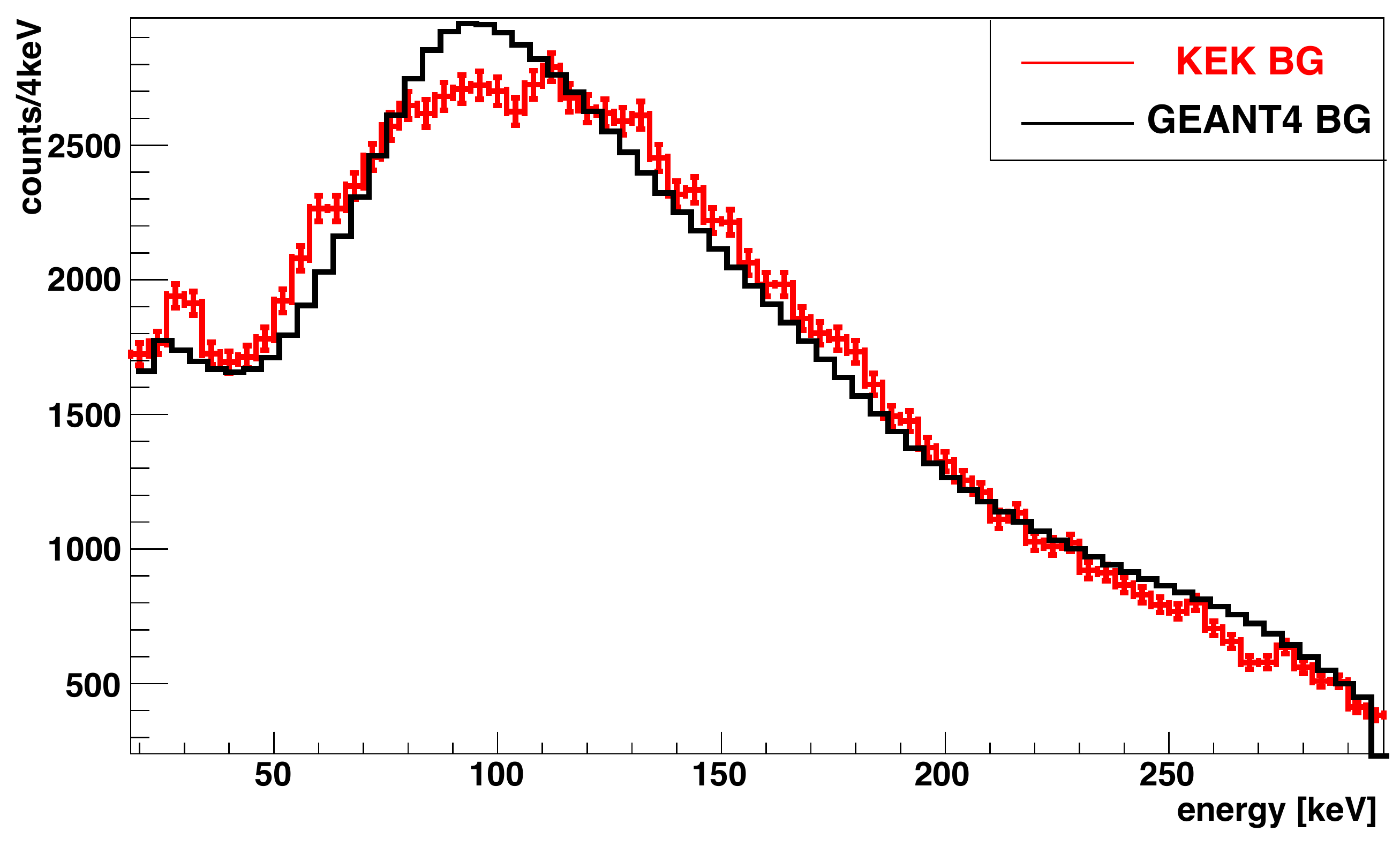}
\end{center}
\caption{The background models for the Al target obtained from the experimental data (KEK BG) and the GEANT4 simulation (GEANT4 BG).}
\label{Al_KEK_GEANT_BG}
\end{figure}

As described above, the cuts applied to the TOF timing and dE/dX energy deposit provide excellent antiproton selection in the original beam (see Fig \ref{Al_fit_KEK}). Thus, the main background is due to the annihilation products of the exotic atom, which can develop an electromagnetic shower in the target and the detector frame. Similarly, most of the antiprotons in the beam were annihilated in the degrader and the annihilation products developed the electromagnetic shower around the detector. Therefore, the background spectrum was estimated with the experimental data with cuts on the TOF timing at TAC1 (between the P0 and P1 counters) and TAC2 (between the P0 and P2 counters) to evaluate the electromagnetic shower generated by the annihilation products. Note that the background spectrum was also modeled with a GEANT4 simulation and both models (KEK BG, GEANT4 BG) are compared in Figure \ref{Al_KEK_GEANT_BG}. They are in good agreement except that the GEANT4 BG model has a slightly narrower peak around 100 keV. This could be due to the imperfection of the complicated physics process on the antiproton annihilation and the subsequent shower development in the simulation. In order to check the robustness to the deviation of the background model, the fitting results with the KEK BG model were compared to the ones with the GEANT4 BG model (see Section \ref{Sec:Al_fit}). Both results show a good agreement with each other and thus we will use the KEK BG model in the analysis below.

\subsection{Atomic X-ray Spectrum}

\begin{figure}[!b]
\begin{center} 
\includegraphics*[width=7.5cm]{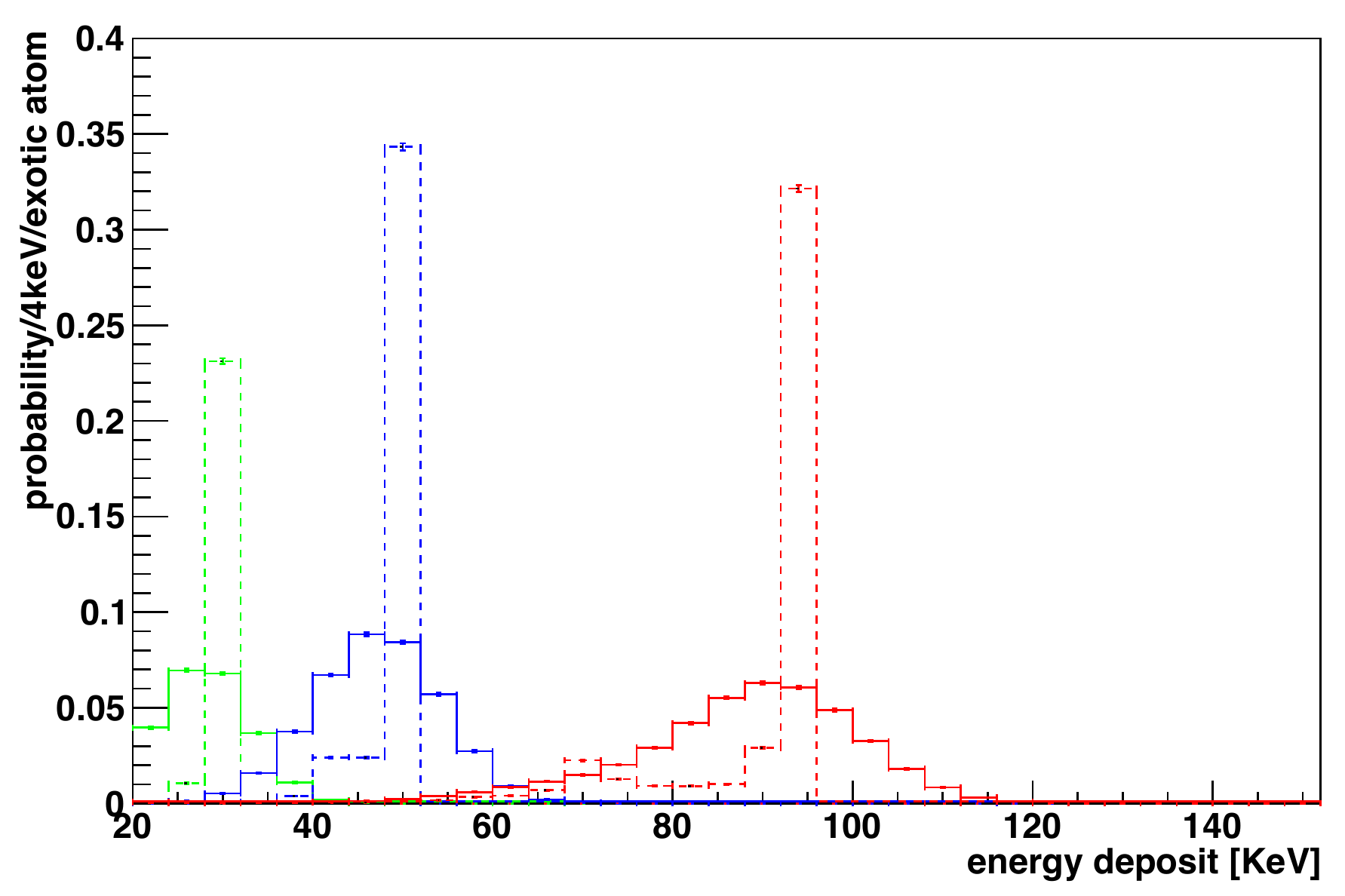}
\end{center}
\caption{Expected energy spectra in the detector for each atomic X-ray with the Al target. The green, blue, and red lines represent simulation results for 30 keV, 50 keV, and 92 keV X-rays, and the solid lines are the spectra with the detector response. It is normalized to the counts per exotic atom with 100\% X-ray yield.}
\label{Al_Xray}
\end{figure}

\begin{figure}[!b]
\begin{center} 
\includegraphics*[width=7.5cm]{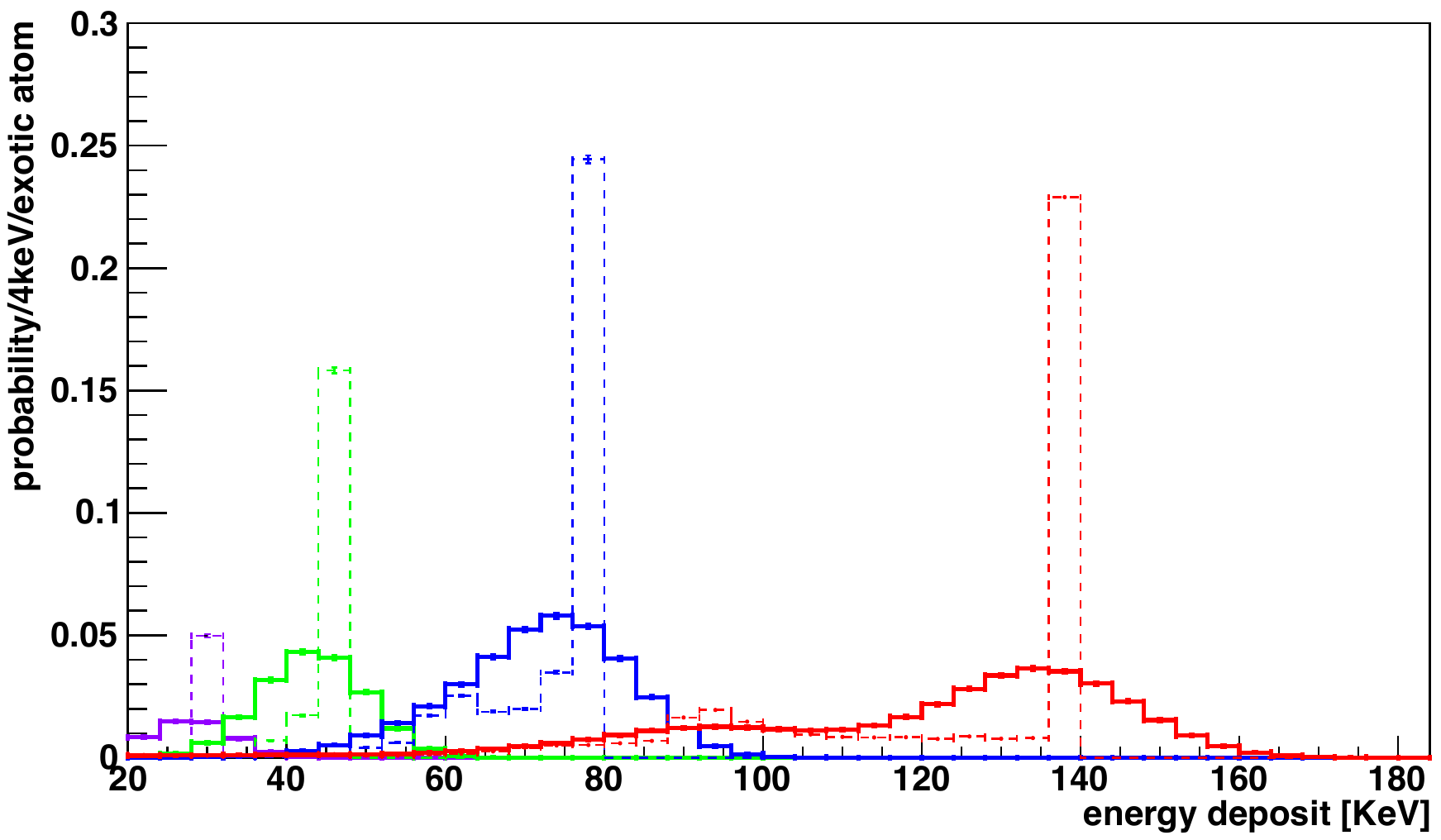}
\end{center}
\caption{Expected energy spectra in the detector for each atomic X-ray with the S target. The purple, green, blue, and red lines represent simulation results for 30 keV, 46 keV, 76 keV and 139 keV X-rays, and the solid lines are the spectra with the detector response. It is normalized to the counts per exotic atom with 100\% X-ray yield.}
\label{S_Xray}
\end{figure}

Since antiprotons were not able to be tracked after hitting the P4 counter, the GEANT4 simulation was used to predict the stopped position of the incoming antiprotons. The simulation was also used to estimate the energy spectrum in the detector for each atomic X-ray, taking into account all the X-ray interactions before reaching the detector. Figure \ref{Al_Xray} shows the expected energy spectra in the detector for each atomic X-ray with the Al target. It is normalized to the counts per exotic atom with 100\% X-ray yield, and the green, blue, and red lines represent 30 keV, 50 keV, and 92 keV X-rays. The dashed lines are the simulation results without the detector response and the solid lines are the spectra with the detector response (7\% FWHM at 1 MeV). Figure \ref{S_Xray} is the same for the S target (30 keV for purple, 46 keV for green, 76 keV for blue and 139 keV for red lines).
 
\subsection{X-ray Yields of Antiprotonic Exotic Atom}

The absolute X-ray yields (probability to emit an atomic X-ray per exotic atom) for antiprotonic exotic atoms, $Y$, were estimated by fitting the data with the background model and the expected energy spectra for each atomic X-ray in the detector as below. 

\begin{eqnarray*}
f_{data} = a_{BG} \cdot f_{BG} + \sum a_{i} \cdot f_{i} 
\end{eqnarray*}
\begin{eqnarray*}
Y_{i} = a_{i} / N_{\bar{p}_{\scriptsize \mbox{stop}}}
\end{eqnarray*}
Here, $a$'s are the fit parameters, $i$ is for the $i$-th atomic X-ray, $f$'s are the spectra in the detector as discussed above, and $N_{\bar{p}_{\scriptsize \mbox{stop}}}$ is the number of stopped antiprotons in the target. The parameter $a_{i}$ also denotes the number of the atomic X-rays emitted from the exotic atom. 

\subsubsection{Al Target}
\label{Sec:Al_fit}

\begin{figure}[!b]
\begin{center} 
\includegraphics*[width=7.5cm]{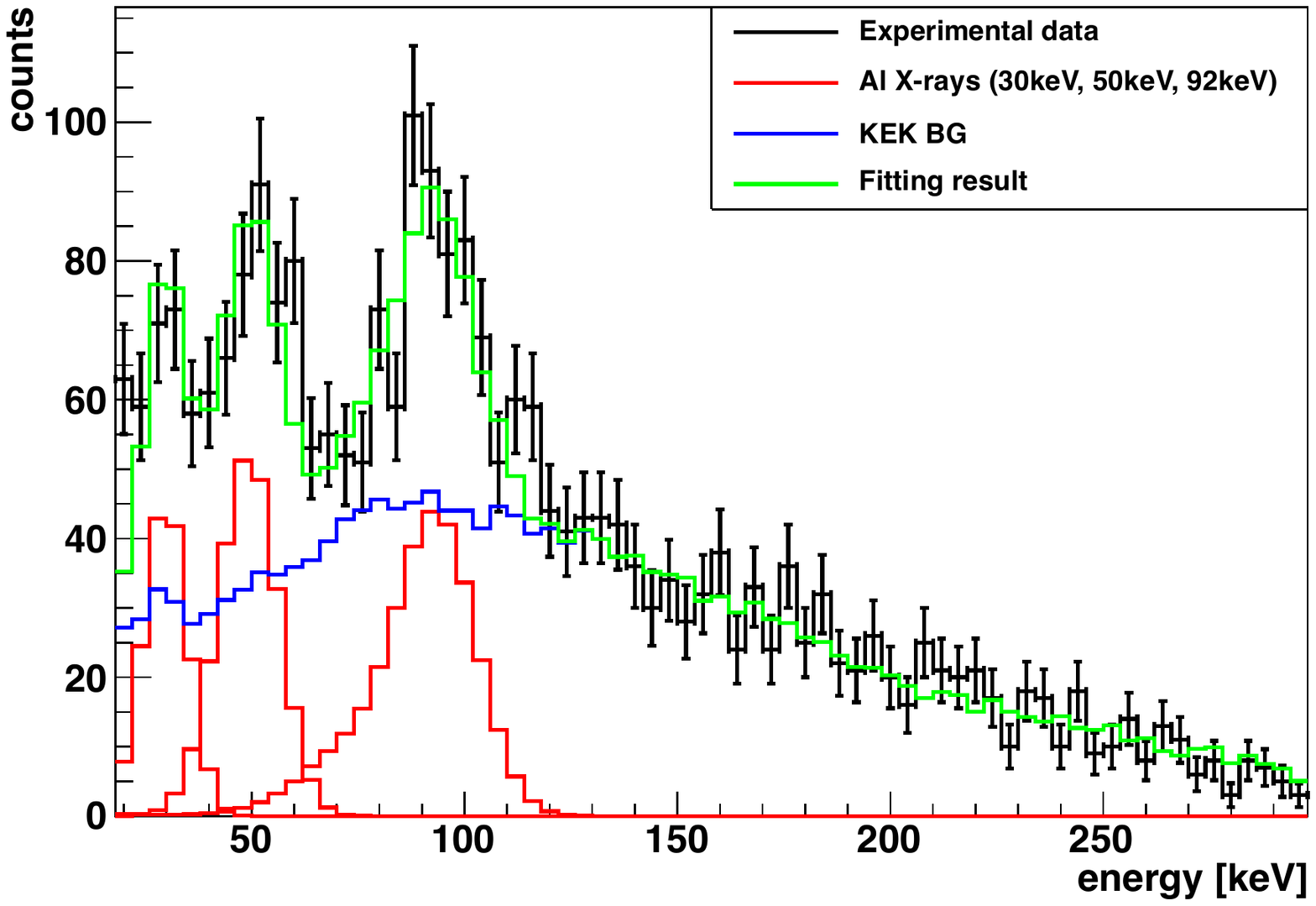}
\end{center}
\caption{The data for the Al target fitted with the background model obtained from the experimental data (blue) and the expected X-ray spectrum for each antiprotonic X-ray (red).} 
\label{Al_fit_KEK}
\end{figure} 

\begin{table}[hbtp]
\centering
\begin{tabular}{c|c|c}
$\bar{p}$-Al & Transition & Yield\\
\hline
92 keV & $5 \rightarrow 4$ & 90\% $\pm$ 13\% \\
\hline
50 keV & $6 \rightarrow 5$ & 76\% $\pm$ 10\%\\
\hline
30 keV & $7 \rightarrow 6$ & 84\% $\pm$ 13\%\\
\hline
reduced-$\chi^2$ & - & 1.07\\
\end{tabular}
\caption{The fitting result for the Al target. }
\label{Yield_fit_Al}
\end{table}

Figure \ref{Al_fit_KEK} shows the fitting result for the Al target. The solid black, blue and red lines represent the experimental data, the background model and the three atomic X-rays respectively and the green solid line is the sum of the background model and the X-rays. Table \ref{Yield_fit_Al} shows the X-ray yields for each atomic X-ray including the fitting error and the systematic uncertainty ($Y \pm \Delta Y$). The systematic uncertainties due to the detector response ($\Delta$FWHM $\sim \pm 1\%$ at 1 MeV) and the offset of the energy calibration ($\pm$ 1 keV), the number of stopped antiproton events and the background model are $\sim$ 7\%, $\sim$ 7\% and $\sim$ 4\% respectively. High absolute yields $\sim$ 80\% were seen for all three transitions and the nuclear absorptions were not seen in the $n = 5 \rightarrow 4$ transition, but were seen in the $n = 4 \rightarrow 3$ transition. 

Note that the fitting results with the GEANT4 BG model are 90\% $\pm$ 13\% for 30 keV, 85\% $\pm$ 11\% for 50 keV and 81\% $\pm$ 12\% for 92 keV (reduced-$\chi^2 \sim$ 1.09), which is consistent with the results obtained with the KEK BG model discussed above.

\subsubsection{S Target}

\begin{figure}[!b]
\begin{center} 
\includegraphics*[width=7.5cm]{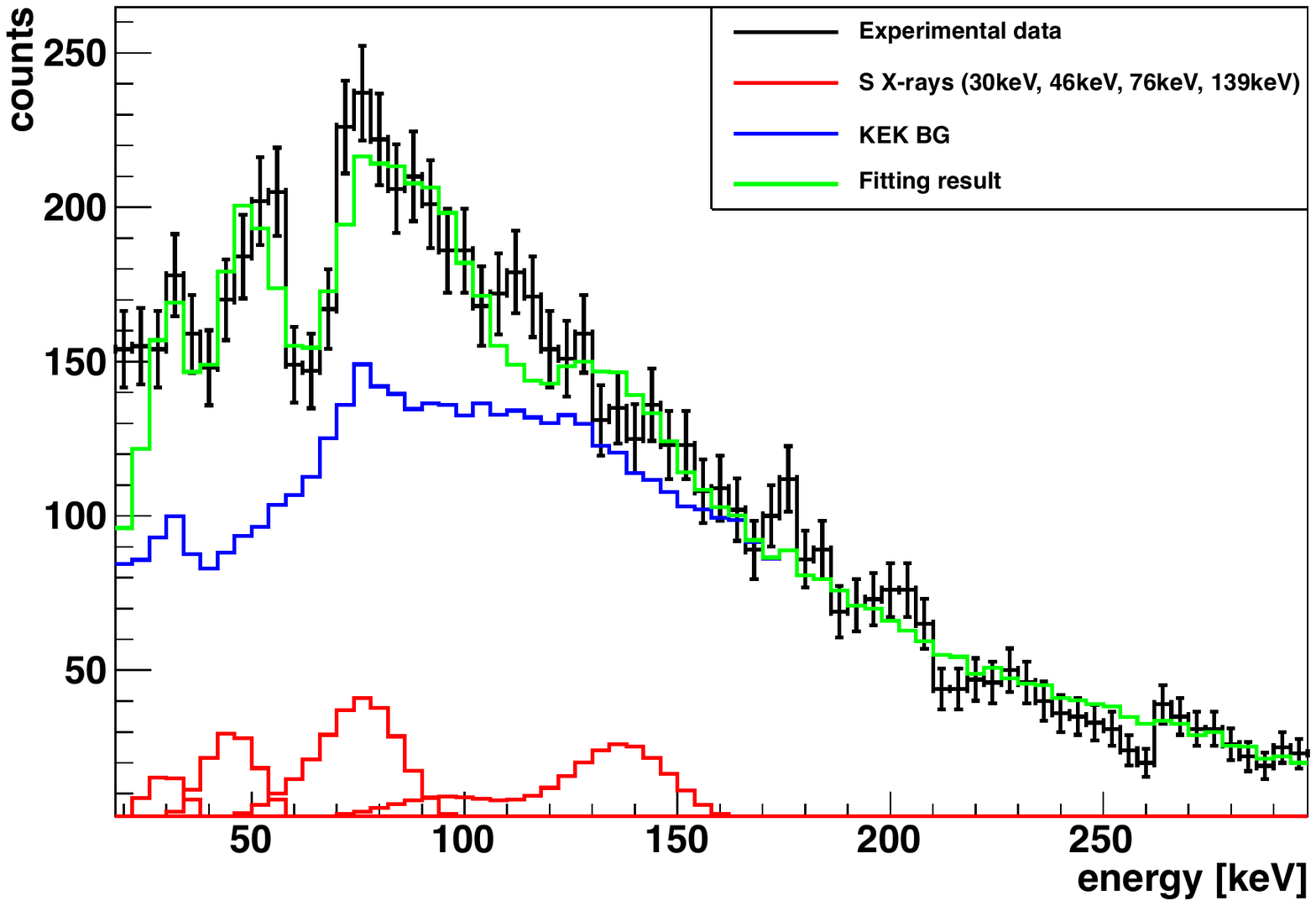}
\end{center}
\caption{The data for the S target fitted with the background model obtained from the experimental data (blue) and the expected X-ray spectrum for each antiprotonic X-ray (red).} 
\label{S_fit_KEK}
\end{figure} 

Since some of the antiprotons may stop in the Al window/frame around the target, seven atomic X-rays (three from the exotic atoms with the Al window/frame and four from the S target) can be produced in the small energy region. Therefore, considering the huge systematic uncertainty, we constrained the three atomic X-rays for the S target, 30 keV ($n = 8 \rightarrow 7$), 46 keV ($n = 7 \rightarrow 6$) and 76 keV ($n = 6 \rightarrow 5$) to have the same absolute yields. This is consistent with theoretical expectation for the S target and as predicted and measured for the Al target. Additionally, the X-ray yields for the antiprotons stopped in the Al window/frame are constrained with the value obtained for the Al target. 
 
 \begin{table}[hbtp]
\centering
\begin{tabular}{c|c|c}
$\bar{p}$-S & Transition & Yield\\
\hline
139 keV & $5 \rightarrow 4$ & 59\% $\pm$ 20\\
 \hline
76 keV & $6 \rightarrow 5$ & 72\% $\pm$ 18\\
 \hline
46 keV & $7 \rightarrow 6$ & 72\% $\pm$ 18\\
\hline
30 keV & $8 \rightarrow 7$ & 72\% $\pm$ 18\\
\hline
reduced-$\chi^2$ & - & 1.00 \\
\end{tabular}
\caption{The fitting result for the S target. }
\label{Yield_fit_S}
\end{table}

Figure \ref{S_fit_KEK} shows the fitted results for the S target. Same as in Fig \ref{Al_fit_KEK}, the solid black, blue and red lines represent the experimental data, the background model and the three atomic X-rays respectively and the green solid line is the sum of the background model and the X-rays. Table \ref{Yield_fit_S} shows the X-ray yields for each atomic X-ray including the fitting error and the systematic uncertainty ($Y \pm \Delta Y$) as discussed above. High absolute yields were also seen in all the transitions except for the $n = 5 \rightarrow 4$ transition, which is due to the nuclear absorption and consistent with the result shown in \cite{Kunselman1973}. 

\subsection{Comparison with Cascade Model}

As discussed in Section \ref{Sec:Cascade}, the cascade model has been developed to estimate the X-ray yields of the exotic atoms at low $n$ states. The yields are mainly determined by the parameter $a$, initial angular momentum distribution, while the last transition rate is strongly depending on $W$, nuclear potential (see Section \ref{Sec:Cascade}). The cascade model with the parameters, $a = 0.16$, $W = 5$ MeV and $\Gamma_{ref} = 10^{14}$ s$^{-1}$ is quite consistent with the experimental data for the Al target as seen in Table \ref{Yield_Al}. Table \ref{Yield_S} shows the X-ray yields of the experimental data and the cascade model for the S target, which is also in good agreement with the experimental data. Parameters used here are, $a = 0.16$, $W = 5$ MeV and $\Gamma_{ref} = 10^{14}$ s$^{-1}$. 

\begin{table}[hbtp]
\centering
\begin{tabular}{c|c|c}
$\bar{p}$-Al & Experiment & Cascade Model\\
\hline
92 keV ($5 \rightarrow 4$) & 90\% $\pm$ 13\% & 78\% \\
\hline
50 keV ($6 \rightarrow 5$) & 76\% $\pm$ 10\% & 84\% \\
\hline
30 keV ($7 \rightarrow 6$) & 84\% $\pm$ 13\% & 71\%\\
\end{tabular}
\caption{The experimental data and the cascade model for X-ray yields of antiprotonic exotic atom with the Al target ($a = 0.16$, $W = 5$ MeV and $\Gamma_{ref} = 10^{14}$ s$^{-1}$).}
\label{Yield_Al}
\end{table}

\begin{table}[hbtp]
\centering
\begin{tabular}{c|c|c}
$\bar{p}$-S & Experiment & Cascade Model\\
\hline
139 keV ($5 \rightarrow 4$) & 59\% $\pm$ 20\% & 50\%\\
 \hline
76 keV ($6 \rightarrow 5$) & 72\% $\pm$ 18\% & 83\%\\
 \hline
46 keV ($7 \rightarrow 6$) & 72\% $\pm$ 18\% & 78\%\\
\hline
30 keV ($8 \rightarrow 7$) & 72\% $\pm$ 18\% & 60\%\\
\end{tabular}
\caption{The experimental data and the cascade model for X-ray yields of antiprotonic exotic atom with the S target ($a = 0.16$, $W = 5$ MeV and $\Gamma_{ref} = 10^{14}$ s$^{-1}$).}
\label{Yield_S}
\end{table}

\subsection{Prediction for Antiprotonic and Antideuteronic Exotic Atom for Si Target}
\label{Sec:dbar_exotic_atom}
The cascade model was extended to the antiprotonic and antideuteronic exotic atom for a Si target and other materials in the GAPS instrument to estimate the antideuteron sensitivity. Since parameters are not strongly correlated with the atomic number, the same parameters are used for the Si (Z = 14) target as used for the Al (Z = 13) and S (Z = 16) targets. Table \ref{Yield_Si} shows the result for the antiprotonic exotic atom with the Si target ($a = 0.16$, $W = 5$ MeV, $\Gamma_{ref} = 10^{14}$ s$^{-1}$).

\begin{table}[hbtp]
\centering
\small\addtolength{\tabcolsep}{-4pt}
\begin{tabular}{c|c|c}
$\bar{p}$-Si & Cascade Model & Ref. in \cite{Mori2002} \\
\hline
106 keV ($ 5 \rightarrow 4 $) & 70\% & 50\%\\
 \hline
 58 keV ($ 6 \rightarrow 5 $) & 84\% & 50\% \\
 \hline
 35 keV ($ 7 \rightarrow 6 $) & 73\% & 50\% \\
\end{tabular}
\caption{Cascade model for X-ray yields of the antiprotonic exotic atom with the Si target ($a = 0.16$, $W = 5$ MeV, $\Gamma_{ref} = 10^{14}$ s$^{-1}$).}
\label{Yield_Si}
\end{table}

The X-ray yields for the antideuteronic exotic atom with a Si target were also estimated by simply changing the optical potential, $W_{\bar{d}} \sim 2W_{\bar{p}}$ = 10 MeV, as shown in Table \ref{Yield_Si_dbar} ($a = 0.16$, $\Gamma_{ref} = 10^{14}$ s$^{-1}$). It was also estimated for higher values of $W$ = 20 MeV, however, the result does not affect the GAPS antideuteron sensitivity since the nuclear capture only takes place at $n = 6$, and the corresponding atomic X-ray energy (112 keV) is too high to be detected in the GAPS detector. The result indicates an increase of GAPS sensitivity to the antideuterons \cite{Aramaki2013} since the X-ray yields for the GAPS experiment were previously assumed to be $\sim$ 50 \% in \cite{Mori2002}. 

\begin{table}[hbtp]
\centering
\small\addtolength{\tabcolsep}{-4pt}
\begin{tabular}{c|c|c|c}
$\bar{d}$-Si & $W$ = 10 MeV & $W$ = 20 MeV & Ref. in \cite{Mori2002} \\
\hline
 112 keV ($ 6 \rightarrow 5 $) & 28\% & 17\% & -\\
\hline
 67 keV ($ 7 \rightarrow 6 $) & 96\% & 94\% & 50\% \\
 \hline
 44 keV ($ 8 \rightarrow 7 $) & 92\% & 93\% & 50\% \\
\hline
 30 keV ($ 9 \rightarrow 8) $ & 80\% & 80\% & 50\% \\
\end{tabular}
\caption{Cascade model for X-ray yields of antideuteronic exotic atom with the Si target ($a = 0.16$, $\Gamma_{ref} = 10^{14}$ s$^{-1}$).}
\label{Yield_Si_dbar}
\end{table}

\section{Conclusion}
\label{Sec:Conclusion}

Absolute X-ray yields for the antiprotonic exotic atom with Al and S targets were measured at KEK, Japan in 2005. We obtained high X-ray yields, $\sim$ 75\%, for both targets at low $n$ states. The nuclear absorption was seen only in the very low $n$ state for the S target. A simple but comprehensive cascade model has been developed to estimate the X-ray yields of the exotic atom. Since it is extendable to any kind of exotic atom (any negatively charged cascading particles with any target materials), the model was evaluated and validated with the experimental data and other models for the antiprotonic and muonic exotic atoms. The model allows us to estimate the X-ray yields of the antiprotonic and antideuteronic exotic atoms formed with any materials in the GAPS instrument and the X-ray yields for antiprotonic and antideuteronic exotic atoms with a Si target were estimated as $\sim$ 80\%. This is higher than previously assumed in \cite{Mori2002}, indicating the increase of the GAPS antideuteron sensitivity. The subsequent GAPS antideuteron sensitivity \cite{Aramaki2013} indicates that the GAPS project has a strong potential to detect antideuterons produced by dark matter. 

\section{Acknowledgments}
We would like to thank J Collins and the electronics shop staff at LLNL for the development and construction of the GAPS electronics, and T Decker, R Hill and G Tajiri for mechanical engineering support. We would also like to thank T Koike for the helpful discussion on the cascade model. We gratefully acknowledge the support of M Ieiri and the KEK staff, and J Jou before and during the accelerator experiments. This work was supported in part by a NASA SR\&T grant, NAG5-5393.





\bibliographystyle{model1-num-names}

\bibliography{refs}






\end{document}